\documentstyle[aps,eqsecnum,preprint]{revtex}
\tighten
\begin{document}

\title{Foundations of self-consistent particle-rotor models and of
self-consistent cranking models}

\author{Abraham Klein\footnote{
Email: {aklein@nucth.physics.upenn.edu}}}
\address{Department of Physics, University of Pennsylvania, 
Philadelphia, PA 19104-6396, USA}

\maketitle 
 
\medskip
\begin{abstract}

The Kerman-Klein formulation of the equations of motion for a nuclear shell
model and its associated variational principle are reviewed briefly.  It is then
applied to the derivation of the self-consistent particle-rotor model and of
the self-consistent cranking model, for both axially symmetric and triaxial
nuclei.  Two derivations of the particle-rotor model are given. One of these is of a form that lends itself to an expansion of the result in powers of the
ratio of single-particle angular momentum to collective angular momentum, that
is essential to reach the cranking limit.  The derivation also
requires a distinct, angular-momentum violating, step. 
The structure of the result implies the possibility
of tilted-axis cranking for the axial case and full three-dimensional 
cranking for the triaxial one.  The final equations remain number conserving.
In an appendix, the Kerman-Klein method is developed in more detail, and the outlines of several algorithms for obtaining solutions of the associated 
non-linear formalism are suggested.

\end{abstract}
\bigskip


\newpage

\section{Introduction}

The aims of this paper are to study anew the foundations of the particle-rotor
model \cite{BM,RS}  and of the cranking model \cite{RS}.  The particle-rotor
model (PRM) was introduced  as an angular momentum-conserving  
phenomenological description of odd deformed nuclei. Because of its relative
ease of application and, on the whole, quite remarkable success, it has been
applied even up to the present (for instance \cite{SMS}), with various alterations of detail, to a myriad of applications,
over a lifetime of more than four and a half decades. 
Among the extensions, we mention in particular that to the description of
triaxial nuclei \cite{MTV,LLR,RagS}, the original model having been formulated for
axially symmetric nuclei. 
 
In one of the textbooks \cite{RS}, p.\ 109, we find,
after a glowing appraisal of the success of the model, the following statement: ``However,
until now a clear-cut microscopic derivation has been missing.'' In fact, a 
microscopic derivation had been given earlier \cite{cpc}, based on the Kerman-Klein (KK) method \cite{kk:1,CKK,kk2,kk3,kk4}. 
The microscopic foundation of the axially symmetric
PRM was studied more recently in \cite{pav4}, starting from a 
semi-microscopic version of the KK approach, and compared in accuracy, for 
several examples of well-deformed nuclei, both with its more accurate progenitor
and with the inherently less accurate cranking approximation.

The cranking model was originally introduced into nuclear physics \cite{In}, 
within the framework of a prescribed single-particle model,
to deal with the enigma presented by the first values encountered for the moments of inertia of deformed nuclei.  An extended version \cite{TV}, the one considered
in most applications until recent years, was based on the self-consistent
mean-field theory of a deformed rotating object.  This early work was designed
primarily to provide formulas for the moment of inertia. 
 
The full range of applicability of the self-consistent cranking model, as well as its limitations, was realized in the so-called cranked shell model (CSM) \cite{BF}, that has been
widely applied to the analysis of band-crossing and other high-spin phenomena.
(For a current list of references, especially reviews, see \cite{F1}.) The formulations under discussion, which apply to axially symmetric nuclei, assume that collective
rotation occurs about a principal axis perpendicular to the symmetry axis.  Such
a formulation is referred to currently as principal-axis cranking (PAC) as opposed
to a recent generalization, called tilted-axis cranking (TAC) 
\cite{F1,F2,F3,F4,F5,}.  In the latter, even in the axial case, the system may
rotate about an axis in a principal plane of the assumed (quadrupole) intrinsic
shape, and for the triaxial case about an arbitrary (dynamically determined) 
direction with respect to the principal axes.

A second aim of the present paper is to establish the relationship of the 
cranking models, including the recent generalized versions, to a microscopic
theory.  The previous literature on this subject is modest in extent.  The standard references are \cite{RBM,BMR}, the major results of which are reproduced
and discussed in \cite{RS}.  Briefly, starting from a formulation
of the microscopic theory by means of generator coordinates, the energy is evaluated approximately as a power series in the angular momentum by a method due to Kamlah \cite{Kam}, valid for large deformations.  When the variational method is applied to the lowest non-trivial approximation of this procedure, it can be shown that the cranking theory
is a solution of the resulting equations.  This is summarized by stating that
cranking is a solution, involving a semi-classical approximation, of the method of variation {\em after} projection as opposed to the exact procedure of variation {\em before} projection.  

To our knowledge, the only other studies of this
subject are those based on the KK method, a brief treatment of the case of rotation
in a plane \cite{KD} that predates the above-cited work and two studies that
postdated them, one again on the problem of rotation in a plane \cite{VK1} and 
the second a restricted study of the triaxial case \cite{VK2}.  (Some discussion of the cranking limit, also based on a variant of the KK method, can be found in
\cite{var2}.)
Up to now we have never presented a full account of the three-dimensional
treatment either for axial or for triaxial nuclei.  Since our methods contain
features distinct from those found in the standard literature \cite{RS},
and in view of the renewed interest in generalized cranking models 
\cite{F1,F2,F3,F4,F5}, the publication of a detailed account is perhaps justified, even at this late date.

The foundations of the study are presented in Sec. II.  We utilize a shell
model Hamiltonian, widely employed for medium and heavy nuclei, with two-particle interactions in which the latter are separated
into two parts clearly distinguished as multipole and pairing forces, 
respectively.  The advantage of such a model is that the (c-number) equations of motion that can be derived from it by the KK method are completely rigorous.  It is a simplifying
feature for the further study to recognize that these equations can be derived from a variational principle that we called the trace variational principle, suggested in our earliest paper \cite{kk:1} and developed more fully
in \cite{var1} and in \cite{var2}.  This variational principle has several noteworthy features:  (i) It is formulated for the many-body problem in the
language of second quantization.  (ii)  The quantities varied are not wave
functions, but rather a suitably chosen set of matrix elements, in our case 
coefficients of fractional parentage (to be discussed at the appropriate point of
Sec.\ II).  (iii) Rather than involving the Rayleigh-Ritz principle for one
state at a time, the functional to be varied is the trace of energy expectation
values over a prescribed space of states.  

[{\footnotesize It turns out that
not all aspects of our formulation are novel.  Thus an incomplete version
of the trace variational principle is to be found in one of the initial series of
papers on matrix mechanics \cite{BJ}, in which the variational parameters are
matrix elements of the coordinates and momenta.  This application to particle 
quantum mechanics was discovered
and developed independently by us in several accounts of which the most recent is
\cite{LKM3}.  A version of the trace variational principle can, furthermore, 
be found is in a classic 
text in mathematical physics \cite{CH}.  Here the formulation is close to standard
Rayleigh-Ritz, in that the quantities varied are wave function.  This formulation
has found its way into the theory of density functionals \cite{Theo} and even
been generalized to include the case where the trace is replaced by a different
weighted diagonal sum \cite{GOK}.  Most recently the trace variational principle for fields has appeared in a quaternion generalization of quantum mechanics \cite{SLA}.}]

The theory is elaborated in Sec.\ II only as far as is required for the remaining body of the text.  Further development is presented in Appendix A, with an eye
to formulating algorithms that can eventually be applied to the study of the 
self-consistent problem posed by the formulation in Sec.\ II. We turn to applications in Sec.\ III, where we derive the self-consistent PRM from the variational principle associated with the KK equations.  (With one possible exception \cite{Brink}, we are unaware of any recent work, other than our own, that has 
examined the foundations of the PRM.)  

The formalism presented in Sec.\ III does not lend itself naturally to a derivation of the self-consistent cranking theory, which should be a limit of 
the self-consistent PRM.  In Sec.\ IV we describe an alternative derivation
of the PRM,
following ideas first advanced briefly in \cite{KD}, that does lead directly to the cranking limit.  The considerations of Secs.\ III and IV apply to axially symmetric nuclei.  Both
treatments are extended to the case of triaxial nuclei in Sec.\ V. Further discussion of results and conclusions are given in Sec.\ VI.

\section{Equations of motion and variational principle}

We choose a shell-model Hamiltonian in the form
\begin{eqnarray}
H &=& h_a a_\alpha^{\dag}a_\alpha +\frac{1}{2}F_{\alpha\gamma
\delta\beta}a_\alpha^{\dag}a_\gamma a_\beta^{\dag}a_\delta
  +\frac{1}{2}G_{\alpha\gamma\beta\delta}a_\alpha^{\dag}a_\gamma^{\dag}
a_\delta a_\beta.  \label{kc.ham}
\end{eqnarray}
In this standard model, the $a_\alpha,a_\alpha^{\dag}$ are the destruction,
creation operators for fermions in the shell-model mode $\alpha=(nljm\tau)$
($\tau$ distinguishing neutrons from protons); $F_{\alpha\gamma\delta\beta}$
describes multipole forces and $G_{\alpha\gamma\delta\beta}$ pairing forces.
In this version, all multipolarities allowed by angular momentum conservation
are included, though in practice we limit ourselves to the lowest few
multipoles of each type.  We shall also consistently use the summation
convention, except when we wish to highlight some set of indices.
With the help of the definitions
\begin{eqnarray}
F_{\alpha\gamma\delta\beta} &=& s_\gamma(j_a m_a j_c -m_c|L M_L)
s_\beta(j_d m_d j_b -m_b|LM_L)F_{acdb}(L), \label{kc.f} \\
G_{\alpha\gamma\delta\beta} &=& (j_a m_a j_c m_c|LM_L)
(j_d m_d j_b m_b|LM_L)G_{acdb}(L),  \label{kc.g} \\
s_\gamma &=& (-1)^{j_c -m_c} =\sqrt{2j_c +1}(j_c m_c j_c -m_c|
00), \label{kc.phase}
\end{eqnarray}
where $(jmj'm'|LM)$ is a Clebsch-Gordon (CG) coefficient,
the operator equations of motion can be obtained in the form
\begin{eqnarray}
[a_\alpha,H] &=& h_a^{\prime}a_\alpha +F_{\alpha\alpha'\beta'\beta}a_{\alpha'} a_\beta^{\dag}
a_{\beta'} +G_{\alpha\alpha'\beta\beta'}a_{\alpha'}^{\dag}
a_{\beta'} a_\beta,  \label{kc.eom1} \\
h_a^{\prime} &=& h_a -\frac{1}{2}F_{abab}\frac{2L+1}{2j_a +1},
\label{kc.sp1} \\
{}[a^{\dag}_{\bar{\alpha}},H] &=& -h_a^{\prime\prime}a_{\bar{\alpha}}^{\dag}
-F_{\beta\beta'\alpha'\bar{\alpha}}
a_{\beta'}^{\dag}a_\beta a_{\alpha'}^{\dag} -G_{\beta\bar{\beta}'\alpha'
\bar{\alpha}}a_{\alpha'} a_\beta^{\dag}a_{\bar{\beta}'}^{\dag}, \label{kc.eom2}\\
h_a^{\prime\prime} &=& h_a^{\prime}+2\frac{2L+1}{2j_a +1}
G_{abab}(L).  \label{kc.sp2}
\end{eqnarray}
Here, for example, $\bar{\alpha}=(j_a, -m_a)$.

To develop a dynamical scheme, we turn to the problem of obtaining equations for the matrix elements of Eqs.~(\ref{kc.eom1}) and (\ref{kc.eom2}).  We
designate a state of interest of an odd nucleus as $|JM\nu\rangle$, where $J$ is the total angular momentum,
$M$ is its $z$ component, and $\nu$ are the remaining quantum numbers necessary for unique specification of the state.  
Neighboring even nuclei are specified, correspondingly, as $|\overline{IMn}\rangle$, referring to a heavier neighbor, and
$|\underline{IMn}\rangle$, referring to a lighter neighbor.
Below we shall then derive equations for the matrix elements, referred
to as CFP (coefficients of fractional parentage) 
\begin{eqnarray}
\langle JM\nu|a_\alpha|\overline{IM_I n}\rangle &=&
V_{JM\nu}(\alpha IM_I n),  \label{kc.cfp1} \\
\langle JM\nu|a^{\dag}_{\bar{\alpha}}|\underline{IM_I n}\rangle
&=& U_{JM\nu}(\alpha IM_I n). \label{kc.cfp2}
\end{eqnarray}
We shall require the full notation when we turn to applications
in the next section.  For the formal developments of this section, we utilize a compressed notation, with 
\begin{equation}
JM\nu\rightarrow i,\;\;\; IM_I n\rightarrow n. \label{kc.sh}
\end{equation}

With new symbols defined and discussed below, we thus obtain
the equations
\begin{eqnarray}
{\cal E}_i V_i(\alpha n) &=& (\epsilon_a^{\prime} - E^{\ast}_{\bar{n}})V_i(\alpha n) +F_{\alpha\alpha'\beta'\beta}[V_{i'}^{\ast}(\beta n')
V_{i'}(\beta' n)]V_i(\alpha' n') \nonumber \\
&& 
+G_{\alpha\bar{\alpha}'\beta\bar{\beta}'}[U_{i'}^{\ast}(\beta' n')
V_{i'}(\beta n)]U_i(\alpha' n'),  \label{kc.eom3} \\
{\cal E}_i U_i(\alpha n) &=& 
(-\epsilon_a^{\prime\prime}-E_{\underline{n}}^{\ast})U_i(\alpha n) 
-F_{\bar{\beta}\bar{\beta}'\bar{\alpha}'\bar{\alpha}}
[U_{i'}^{\ast}(\beta n')U_{i'}(\alpha' n)]U_i(\beta' n')
\nonumber \\
&& +G_{\bar{\alpha}\alpha'\bar{\beta}'\beta}
[V_{i'}^{\ast}(\beta n')U_{i'}(\beta' n)]V_i(\alpha' n'). 
\label{kc.eom4}
\end{eqnarray}
In the definitions that follow, we understand that $E_i$
is the energy of the state $|i\rangle$ and that $E_{\bar{n}}$
and $E_{\underline{n}}$ are, correspondingly the energies of the 
neighboring even states, with the subscript $0$ standing either for the
ground state, or for the lowest energy state considered, which for
conciseness we shall continue to refer to as the ground state.  
We thus encounter the quantities
\begin{eqnarray}
{\cal E}_i &=& -E_i +\frac{1}{2}(E_{\bar{0}}+E_{\underline{0}}),
\label{kc.def1} \\
\epsilon_a^{\prime} &=& h_a^{\prime} -\lambda, \label{kc.def2}\\
\lambda &=& \frac{1}{2}(E_{\bar{0}}-E_{\underline{0}}), 
\label{kc.def3}  \\
E_n^{\ast} &=& E_n -E_0.  \label{kc.def4} 
\end{eqnarray}

The physical significance of the quantities defined in 
Eqs.~(\ref{kc.def1})--(\ref{kc.def4}) is evident.  ${\cal E}_i$ 
are the negatives
of the energies of the odd nucleus relative to the ground-state energies
of its even neighbors, $\epsilon_a$, variously primed, are single-particle
energies measured relative to the chemical potential $\lambda$, and $E_n^{\ast}$ are excitation energies of the appropriate even nuclei.
Finally in achieving the form of Eq.~(\ref{kc.eom4}), we have assumed that $F$ and $G$ are real.  Given the Hamiltonian (\ref{kc.ham}), Eqs.~ (\ref{kc.eom3})
and (\ref{kc.eom4}) are an exact set of consequences that define a non-linear
eigenvalue problem with eigenvalue ${\cal E}_i$.  The elements on the right hand sides of these equations define an effective Hamiltonian that will be discussed
in considerable further detail in the course of this work.

We display next a functional, ${\cal F}$, whose vanishing first variations yield the equations of motion, namely,
\begin{eqnarray}
{\cal F} &=& \epsilon_a^{\prime}|V_i(\alpha n)|^2 -\epsilon_a^{\prime\prime}|U_i(\alpha n)|^2  \nonumber \\
&& \frac{1}{2}F_{\alpha\alpha'\beta'\beta}[V_{i'}^{\ast}(\beta n')
V_{i'}(\beta' n)][V_i^{\ast}(\alpha n)V_i(\alpha' n')] \nonumber\\
&& +G_{\alpha\bar{\alpha}'\beta'\bar{\beta}}[U_{i'}^{\ast}
(\beta n')V_{i'}(\beta' n)][V_i^{\ast}(\alpha n)
U_i(\alpha' n')] \nonumber \\
&& -\frac{1}{2}F_{\bar{\beta}\bar{\alpha}'\bar{\beta}'\bar{\alpha}}
[U_{i'}^{\ast}(\beta n')U_{i'}(\beta' n)][U_i^{\ast}(\alpha n)
U_i(\alpha' n') \nonumber \\
&& -{\cal E}_i[|V_i(\alpha n)|^2 + U_i(\alpha n)|^2] 
\nonumber\\
&& -E_{\bar{n}}^{\ast} |V_i(\alpha n)|^2
   -E_{\underline{n}}^{\ast} |U_i(\alpha n)|^2 \label{kc.var1}\\
&& \equiv {\cal G} -{\cal E}_i[|V_i(\alpha n)|^2 + |U_i(\alpha n)|^2].
\label{kc.var1a}
\end{eqnarray}
One verifies that the equations of motion (\ref{kc.eom3}) and 
(\ref{kc.eom4}) emerge, respectively, from the requirements
\begin{equation}
\frac{\delta {\cal F}}{\delta V_i^{\ast}(\alpha n)}
= \frac{\delta{\cal F}}{\delta U_i^{\ast}(\alpha n)} =0.
\label{kc.var2}
\end{equation}

It is natural to inquire at this point if the functional
${\cal F}$ has any simple physical significance, in particular,
if it is related to a Rayleigh-Ritz principle.  To answer this
question, we evaluate the sum 
\begin{equation}
\rm{Tr}(\bar{H} +\underline{H})=\sum_n [\langle\bar{n}|H|\bar{n}
\rangle + \langle\underline{n}|H|\underline{n}\rangle]. 
\label{kc.trace}
\end{equation}
The evaluation of this sum with the aim of eventually recognizing the relevant pieces of ${\cal F}$ requires, in addition to the standard tool of completeness, some algebraic
rearrangement of the trace involving the lighter system, just as 
was necessary in the equations of motion.  We then find that the interaction terms match exactly those in Eq.~(\ref{kc.var1}),
but that the single particle terms do not.  Instead we find
\begin{eqnarray}
h_a^{\prime} &\rightarrow& h_a \equiv \bar{h}_a, \label{kc.sp3}\\
h_a^{\prime\prime\prime} &\rightarrow& h_a +2 \frac{2L+1}{2j_a+1}
G_{acac}(L) +\sqrt{\frac{2j_b+1}{2j_a+1}}F_{aabb}(0)
\equiv \underline{h}_a.  \label{kc.sp4}
\end{eqnarray}

We are thus tempted to replace the functional ${\cal F}$, as basis for the theory, by a functional that contains the new single-particle
energies.  We do not make this change because it
destroys the simple physical significance of the Lagrange multiplier terms in Eq.~(\ref{kc.var1}) to which we next turn our attention.  In practice, these extra single-particle terms are often ignored anyway.

We consider then the Lagrange-multiplier terms that appear
in Eq.~(\ref{kc.var1}). The relevant question concerns the constraints that have been imposed on the variations.  Since
\begin{equation}
\sum_{i\alpha}|V_i(\alpha n)|^2 =\sum_{\alpha}\langle \bar{n}|
a_\alpha^{\dag}a_\alpha|\bar{n}\rangle = \langle\bar{n}|\hat{N}|
\bar{n}\rangle,   \label{kc.numc}
\end{equation}
where $\hat{N}$ is the number operator, we see that the excitation energies 
$E^{\ast}_{\bar{n}}$ enter as Lagrange multipliers for the 
conservation of nucleons in the heavier even nucleus.  Similarly
the term involving the sum over the $|U_i(\alpha n)|^2$ expresses
(to an additive constant) the conservation of nucleons in the 
lighter system.  Finally, we show that the eigenvalue ${\cal E}_i$ is (no surprise here) a Lagrange multiplier for an appropriate normalization condition. To see this we take the 
matrix element in the state $|i\rangle$ of the summed anticommutator,
\begin{equation}
\sum_\alpha\{a_\alpha,a_\alpha^{\dag}\}=\Omega,\;\;\;\;
\Omega=\sum_{j_a}(2j_a+1)= \sum_a \Omega_a,   \label{kc.antic}
\end{equation}
and thus find
\begin{equation}
\frac{1}{\Omega}\sum_{\alpha n}[|V_i(\alpha n)|^2 
+|U_i(\alpha n)|^2] = 1. \label{kc.norm}
\end{equation}
Orthogonality constraints on the solutions need not be imposed,
since they follow directly from the equations of motion. 

There is more to the story, however.  We must note that Eq.~(\ref{kc.norm})
is only a sum of required normalization conditions.  From the summed anticommutator for each level,
\begin{equation}
\sum_{m_a} \{a_\alpha,a_\alpha^{\dag}\} = \Omega_a, \label{kc.antica},
\end{equation}
we have 
\begin{equation}
\frac{1}{\Omega_a}\sum_{m_a n}[|V_i(\alpha n)|^2 + U_i(\alpha n)|^2] =1.
\label{kc.norma}
\end{equation}
If Eqs.~(\ref{kc.eom3}) and (\ref{kc.eom4}) described a linear eigenvalue 
problem, it would be impossible to impose the additional normalization 
conditions represented by Eq.~(\ref{kc.norma}).  For the general non-linear
problem, there is no {\em a priori} inconsistency;  the satisfaction of these
conditions will be a part of any fully satisfactory algorithm.  The form of the normalization condition (\ref{kc.norma}) suggests, furthermore, that it may 
be both useful and natural to rescale the CFP,
\begin{eqnarray}
V_i(\alpha n)& =& \sqrt{2j_a +1}v_i(\alpha n), \label{littlev} \\
U_i(\alpha n)& = & \sqrt{2j_a +1}u_i(\alpha n). \label{littleu}
\end{eqnarray}

There is considerably more to the formal theory than what has been presented thus far.  However, we have all the tools needed for the further development in the text and thus relegate the additional theoretical considerations to Appendix A.

\section{Derivation of particle-rotor model and its cranking limit:
axially symmetric case}

As a first illustration of the formalism presented in the previous section, we
assume that the even (core) nuclei are in a single axially-symmetric band
$|IM_IK\rangle$, where $K$ is the component of the angular momentum along the
figure axis.  There are at least two cases where it makes some physical sense
to isolate a single $K$ value, where it is the ground-state band with $K=0$,
or where the band has a large $K$-value and we are dealing with an isomeric
state.

We first use rotational invariance to study the structure of the 
amplitudes $V$ and $U$ defined in Eqs.~(\ref{kc.cfp1}) and (\ref{kc.cfp2}),
respectively.  
For this purpose we introduce a complete set of states
$|R\rangle$ localized in the Euler angles, $R=(\alpha\beta\gamma)$, where
$\alpha,\,\beta$ are the usual polar and azimuthal angles, respectively,
and write
\begin{eqnarray}
|IM_I K\rangle &=& \int dR\, |R\rangle\langle R|IM_I K\rangle \nonumber \\
&=& \left(\frac{2I+1}{8\pi^2}\right)^\frac{1}{2} \int dR \,
|R\rangle D^{(I)}_{M_IK}(R). 
\label{kc.Wigner}    \end{eqnarray}
The identification
of a scalar product of {\em many-body states} with the Wigner $D$ function is not a trivial statement, but is rather an essential 
element in the definition of the model to be studied.  In fact,
the designation $|R\rangle$ for the many-body state is  insufficiently detailed and is made more explicit by the 
statement 
\begin{equation}
|R\rangle = U(R)|\hat{0}K\rangle,  \label{kc.intr1}
\end{equation}
where $|\hat{0}K\rangle$ is an axially symmetric intrinsic
state spinning with angular momentum $K$ about its symmetry
axis, and $U(R)$ is the unitary rotation operator in the many-body
space defined by the Euler angles that specify the rotation 
$R$.  For such a state, we thus note the relation, with
$(\alpha\beta\gamma)=(\hat{n}\gamma)$
\begin{equation}
U(\hat{n}\gamma)|\hat{0}K\rangle\exp(-iK\gamma) = U(\hat{n}0)
      |0K\rangle.    \label{kc.intr2}
\end{equation}  
The introduction of strictly localized states is, of course, an idealization 
that ignores the reality of band termination, but it is a standard 
approximation for well-deformed nuclei.  

[{\footnotesize The previous discussion and that which follows does not take into account ${\cal R}$ invariance, the invariance of the quadrupole shape under a rotation of $\pi$ about a principal axis.  To include this symmetry in the discussion, we replace the state $|IM_I K\rangle$ by an eigenfunction of
${\cal R}$,
\begin{equation}
|IM_I K] =\frac{1}{2}\{|IM_I K\rangle +(-1)^{I+K}|IM_I -K\rangle\}.
\end{equation}
We then imitate the arguments starting on p.\ 8 of \cite{BM}.  The task is to sort and collect the extra terms that appear both in the equations of motion and 
in formulas for one and two-particle observables. }]

When Eq.~(\ref{kc.Wigner}) is substituted into the definition
(\ref{kc.cfp1}) of $V$,
and use is made of the definitions to be given below, we are 
thereby led to the study of an amplitude such as
\begin{eqnarray}
\langle JM\nu|a_\alpha |R\rangle &=& \langle JM\nu|U(R)U^{-1}(R)
a_\alpha U(R)|\hat{0}K\rangle \nonumber \\
&=&\sum_{M'} \langle JM\nu|U|JM'\nu\rangle
\langle JM'\nu|U^{-1}a_\alpha U|\hat{0}K\rangle   \nonumber \\
&=&\sum_{M'\kappa_a} D^{(J)\ast}_{MM'}(R)D^{(j_a)\ast}_{m_a\kappa_a }(R)
\chi_{JM'\nu} (j_a\kappa_a,K)(-1)^{j_a +\kappa_a},    \label{kkp4.chi1}
\end{eqnarray}
where the previous manipulations have utilized the following relations 
and definitions (of which the first two are standard):
\begin{eqnarray}
\langle JM|U(R)|JM'\rangle &=& D^{(J)\ast}_{MM'}(R), \label{kkp4.wigner} \\
U^{-1}(R)a_{jm}U(R) &=& \sum_{\kappa}a_{j\kappa}D^{(j)\ast}_{m\kappa}(R), 
\label{kkp4.tensor} \\
\langle JM\nu|a_{jm}|\hat{0}\rangle&\equiv&(-1)^{j+m}\chi_{JM\nu}(jm,K)
\label{kkp4.chi2}
 \end{eqnarray}
The phase in (\ref{kkp4.chi2}) has been introduced for algebraic convenience.

With the help of the integral of a product of three $D$ functions
(\cite{Edmonds}, Eq.~(4.6.2)) and the application of standard symmetry properties of CG coefficients (Eqs.~(3.5.15) and (3.5.16) of the same reference), we find for the CFP defined in Eq.\ (\ref{kc.cfp1}),
\begin{eqnarray}
V_{JM\nu} (\alpha IM_I K) &=& \sum_{\kappa_a}\sqrt{\frac{8\pi^2}{2j_a +1}}
(-1)^{J-M}   \nonumber \\
&& \times (IM_I J-M|j_a m_a)   
(JK-\kappa_a j_a\kappa_a|IK) \nonumber \\
&& \times(-1)^{j_a +\kappa_a}
\chi_{JK-\kappa_a\nu} (j_a\kappa_a,K) . \label{kkp4.xvee}
\end{eqnarray}
A similar analysis carried out for the amplitude $U$ defined in
(\ref{kc.cfp2}) yields the result
\begin{eqnarray}
U_{JM\nu}(\alpha IM_I K) &=&\sum_{\kappa_a} \sqrt{\frac{8\pi^2}{2j_a +1}}
\nonumber \\
&& \times (-1)^{J-M +j_a -\kappa_a +j_a +m_a}(IM_I J-M|j_a m_a) \nonumber \\
&&\times(JK-\kappa_a j_a \kappa_a|IK) 
\phi_{JK-\kappa_a\nu}(j_a\kappa_a,K),   \label{kkp4.xyou}  \\ 
\phi_{JM\nu} (j_a\kappa_a,K) &=& \langle JM\nu|a^{\dag}_{j_a -\kappa_a}
|\hat{0}K\rangle.   \label{kkp4.deffi} 
\end{eqnarray}

It is most succinct to base further discussion on the 
variational principle (\ref{kc.var1}).  
We evaluate this expression when the core collective
states are restricted to the members of a single band of an axial rotor,
and the states of the odd nucleus are any states that can arise from the coupling.
Returning to a full nomenclature, this calls for the identifications
\begin{eqnarray}
&&\bar{n} \rightarrow \overline{IM_I K}, \;\;\; \underline{n} \rightarrow \underline{IM_I K},   \nonumber \\
&& i\rightarrow  JM\nu.     \label{kkp9.4}
\end{eqnarray}
We are assuming here that there are corresponding bands in the two even nuclei
that couple to the given odd nucleus.  In the following we shall also suppress
the bar and underline in the CFP, understanding them from context, but continue to emphasize this distinction in the excitation energies $E^*$.

We consider first the evaluation of all terms in the variational 
principle depending on 
$|V_i(\alpha n)|^2$, which includes some of the single-particle terms and
some of the Lagrange-multiplier terms.   Writing $\bar{E}^*(IK)$ for 
$E^*_{\bar{n}}$, we consider in particular the combination
\begin{equation}
-\sum ({\cal E}_{J\nu} + \bar{E}^*(IK))|V_{JM\nu}(\alpha IM_I K)|^2.
\label{kkp9.5}
\end{equation}
For the evaluation of the CFP $V$, we utilize Eq.~(\ref{kkp4.xvee}),
renormalized, however, by a factor of $(1/\sqrt{8\pi^2})$ in order that the reciprocal of such factors do not appear in the answer, i.\ e.,
we rescale $\chi$ by this factor.  For the evaluation
of expressions involving the CFP $U$, we shall utilize Eq.~(\ref{kkp4.xyou})
similarly renormalized.  This rescaling will be understood throughout the 
remainder of this paper.

The rest of this section consists of a relatively detailed account of the 
evaluation of the variational sum in the ``intrinsic'' system.  Subsequent variation will led to the self-consistent version of the strong coupling PRM.
Toward this end, as part of the definition of an axial rotor, we assume that, equally for the barred and underlined quantities,
\begin{eqnarray}
E^*(IK) &\rightarrow & E^*(\vec{I}^2 - K^2)= E^*(\hat{I}_1^2+\hat{I}_2^2
+\hat{I}_3^2 -K^2), \label{kkp9.7}
\end{eqnarray}
where we have introduced intrinsic components of the angular momentum.
The arrow indicates the replacement of an eigenvalue by an operator.  This
is done by making use of the appropriate one of the CG coefficients, understood as
a scalar product, that appear in Eq.~(\ref{kc.cfp1}), as follows
\begin{equation}
E^*(IK)(JK-\kappa_a j_a\kappa_a|IK)=(JK-\kappa_a j_a\kappa_a|
E^*(\vec{I}^2 -K^2)|IK).   \label{kkp9.100}
\end{equation}
Furthermore, in each term of the sum $I$ is coupled with some $j_a$ to a value
of $J$.  From the structure of the CFP, it follows that we may replace
$\vec{I}$ by $\vec{J} +\vec{j}_a$ in Eq.~(\ref{kkp9.100}) and write
(with $j_a\rightarrow j)$,
\begin{eqnarray}
\bar{E^*}(\vec{I}^2-K^2) &=& \bar{E^*}[(\vec{J}+\vec{j})^2 -K^2]
\nonumber\\
&=& \bar{E^*}(\vec{J}^2 -K^2) +\frac{\partial\bar{E^*}}{\partial \hat{J}_i}
\hat{j}_i +\frac{1}{2}\frac{\partial^2\bar{E}^*}{\partial\hat{J}_i^2}
+ ... \;\;.   \label{kkp9.8}
\end{eqnarray}
It is not necessary for 
these considerations that $E(IK)$ have the simple form of a rotor spectrum, only
that it be a function as indicated.   
The first term of
Eq.~(\ref{kkp9.8}) may be replaced by an eigenvalue
$\bar{E}^*[J(J+1)-K^2)]$, and the second term, the
Coriolis coupling will be evaluated below. The third and possible higher order
terms will not be studied here, but will be included in applications.  Other 
than the Coriolis coupling and higher-order terms,
the contribution from Eq.~(\ref{kkp9.5}) and of the 
remaining single-particle terms takes the form
\begin{eqnarray}
&&\sum_{J\nu j_a\kappa_a}(\epsilon_a^{\prime}-\bar{\varepsilon}_{J\nu})
|\chi_{JK-\kappa_a\nu}(j_a\kappa_a,K)|^2 ,  \label{kkp9.8a}\\
&& \bar{\varepsilon}_{J\nu} ={\cal E}_{J\nu}-\bar{E}^*[J(J+1)-K^2]. \label{defep}
\end{eqnarray}

For the further evaluation we can write in full generality
\begin{equation}
\frac{\partial E^*}{\partial J_i} \equiv f(\vec{J}^2-K^2)J_i \label{kkp9.9},
\end{equation}
where in the simplest case $f$ is just the reciprocal of the 
moment of inertia.
Using the matrix elements of the raising and lowering operators, and collecting terms, we find, in connection with the second term of 
Eq.~(\ref{kkp9.8}), the Coriolis coupling term, the following sum to evaluate
\begin{eqnarray}
&&-\frac{1}{2}\sum_{M_I m_a MIJ\kappa_a\kappa'_{a}}\chi_{JK-\kappa_a\nu}
(j_a\kappa_a,K)
\chi^{\ast}_{JK-\kappa'_a\nu}(j_a\kappa'_{a},K)
(2j_a +1)^{-1}\nonumber\\
&&\times(-1)^{2j_a +\kappa_a +\kappa'_{a}}
|(IM_I J-M|j_a m_a)|^2 (JK-\kappa'_{a} j_{a}\kappa'_{a}|IK) \nonumber \\
&&\times\bar{f}(JK) [\sqrt{(J-K+\kappa_a)(J+K-\kappa_a +1)} \nonumber \\
&&\sqrt{(j_a +\kappa_a)(j_a -\kappa_a +1)} 
(JK-\kappa_a+1 j_a\kappa_a-1|IK) \nonumber \\
&& +\sqrt{(J+K-\kappa_a)(J-K+\kappa_a+1)}\sqrt{(j_a -\kappa_a)(j_a +\kappa_a
+1)} \nonumber\\
&&\times(JK-\kappa_a -1j_a\kappa_a +1|IK)  
+2(K-\kappa_a)\kappa_a(JK-\kappa_a j_a \kappa_a|IK)].   \label{kkp9.10}
\end{eqnarray}
With the help of the standard orthonormalization conditions for 
CG coefficients this reduces to the final result for the 
term under study,
\begin{eqnarray}
&&-\frac{1}{2}\sum_{J\nu j_a\kappa_a}\chi^{\ast}_{JK-\kappa_a\nu}(j_a\kappa_a,K) \bar{f}(JK)    \nonumber \\
&&\times[\sqrt{(J+K-\kappa_a)(J-K+\kappa_a +1)}\sqrt{(j_a -\kappa_a)(j_a +\kappa_a+1)}  \nonumber \\
&& \times\chi_{JK-\kappa_a -1\nu}(j_a\kappa_a +1,K) \nonumber \\
&&+  \sqrt{(J-K+\kappa_a)(J+K-\kappa_a +1)}\sqrt{(j_a +\kappa_a)(j_a -\kappa_a+1)}\nonumber \\
&& \times \chi_{JK-\kappa_a +1\nu}(j_a\kappa_a -1,K)  \nonumber \\
&& +2(K-\kappa_a)\kappa_a\chi_{JK-\kappa_a j_a\kappa_a}(j_a\kappa_a,K)].
\label{kkp9.11}
\end{eqnarray}

We outline briefly the corresponding calculation of the 
single-particle terms
associated with the $U$ coefficients.  Only the following change is necessary: 
All barred energies associated with the heavier of
the two neighboring even nuclei are replaced by underlined energies associated with the lighter of the two neighbors. 
For the terms corresponding to Eq.~\ref{kkp9.8a}), we find
\begin{equation}
-\sum_{J\nu j_a\kappa_a}(\epsilon_a^{\prime\prime}+ \underline{\varepsilon}_{J\nu})
|\phi_{JK-\kappa_a\nu}(j_a\kappa_a,K)|^2 .  \label{kkp9.8b}
\end{equation}
For the Coriolis coupling term, we find
\begin{eqnarray}
&&\frac{1}{2}\sum_{J\nu j_a\kappa_a}\phi^{\ast}_{JK-\kappa_a\nu}(j_a\kappa_a,K)
\underline{f}(JK)  \nonumber \\
&&\times[\sqrt{(J+K-\kappa_a)(J-K+\kappa_a +1)}\sqrt{(j_a -\kappa_a)(j_a +\kappa_a
+1)}  \nonumber \\
&&\times \phi_{JK-\kappa_a -1\nu}(j_a\kappa_a +1,K) \nonumber \\
&&+  \sqrt{(J-K+\kappa_a)(J+K-\kappa_a +1)}\sqrt{(j_a +\kappa_a)(j_a -\kappa_a+1)}\nonumber \\
&& \times \phi_{JK-\kappa_a +1\nu}(j_a\kappa_a -1,K)
+(K-\kappa_a)\kappa_a\phi_{JK-\kappa_a\nu}(j_a\kappa_a,K) ].  \label{kkp9.11b}
\end{eqnarray}

We turn to the contribution of the interaction terms.
We evaluate a typical quartic contribution to 
Eq.~(\ref{kc.var2}), for example 
\begin{eqnarray}
&&\sum_\chi \equiv \frac{1}{2}F_{\alpha\alpha'\beta'\beta}V_{JM\nu}(\beta' IM_IK)
V^{\ast}_{JM\nu}(\beta I'M_{I'}K) \nonumber \\
&&\times V_{J'M'\nu'}(\alpha' I'M_{I'}K)
V^{\ast}_{J'M'\nu'}(\alpha IM_IK) \nonumber \\
&&= F_{aa'b'b}(L)(j_a m_a j_{a'}-m_{a'}|Lm_a -m_{a'})
(j_{b'}m_{b'}j_b -m_b|Lm_{b'}-m_b )\nonumber \\
&&\times(-1)^{j_{a'}-m_{a'}+j_b -m_b}(-1)^{j_a +\kappa_a +j_{a'}+\kappa_{a'}+j_b +\kappa_b +j_{b'} +\kappa_{b'}}  \nonumber \\
&&\times[(2j_a +1)(2j_{a'}+1)(2j_b +1)(2j_{b'}+1)]^{-\frac{1}{2}}
\nonumber \\
&&\times \chi_{JK-\kappa_{b'}\nu}(j_{b'}\kappa_{b'})
\chi^{\ast}_{JK-\kappa_b\nu}(j_b\kappa_b)\chi_{J'K-\kappa_{a'}\nu'}(j_{a'}
\kappa_{a'}) \chi^{\ast}_{J'K-\kappa_a\nu'}(j_a\kappa_a) \nonumber \\
&&\times (IM_IJm_{b'}-M_I|j_{b'}m_{b'})(I'M_{I'}Jm_b -M_{I'}|j_b m_b) \nonumber \\
&&\times(I'M_{I'}J'm_{a'}-M_{I'}|j_{a'}m_{a'})(IM_IJ'm_a -M_I|j_a m_a) \nonumber \\
&&\times (JK-\kappa_{b'}j_{b'}\kappa_{b'}|IK)(JK-\kappa_b j_b\kappa_b|I'K)
\nonumber \\
&&\times(J'K-\kappa_{a'}j_{a'}\kappa_{a'}|I'K)(J'K-\kappa_a j_a\kappa_a|IK).
\label{kkp9.13}
\end{eqnarray}
To obtain this form from the corresponding term of Eq.~(\ref{kc.var1}) we have
utilized Eqs.~(\ref{kkp4.xvee}) and (\ref{kc.f}). 
From angular momentum conservation, we have the relations                
\begin{equation}
m_{b'}-M_I=m_b -M_{I'}, \;\;m_{a'}-M_{I'} =m_a -M_I. \label{kkp9.15}
\end{equation}

The next step is to perform the sums over the magnetic quantum numbers
$m_a$ and $m_b$.  For this purpose, we use Eq.~(6.6.27)  of \cite{Edmonds} 
with assists from Eqs.~(3.5.14)-(3.5.17) of the same reference.
At this point we obtain the sum (henceforth we simplify the arguments of the 
single-particle amplitudes, $j_a\kappa_a,K\rightarrow a$, etc.)
\begin{eqnarray}
\sum_\chi &=&\frac{1}{2}F_{aa'b'b}(-1)^{j_a+\kappa_a+\kappa_{a'}+j_{b'}
+\kappa_{b'}+\kappa_b +J+J'}(2L+1) \nonumber \\
&&\times\chi_{JK-\kappa{b'}\nu}(b')\chi^*_{jK-\kappa_b\nu}(b)
\chi_{J'K-\kappa_{a'}\nu'}(a')\chi^*_{J'K-\kappa_a\nu'}(a) \nonumber \\
&&\times (JK-\kappa_{b'}j_{b'}\kappa_{b'}|IK)(JK-\kappa_b j_b\kappa_b|I'K)   \nonumber\\
&&\times (J'K-\kappa_{a'}j_{a'}\kappa_{a'}|I'K) (J'K-\kappa_a j_a\kappa_a|IK)
\nonumber \\
&&\times\left\{\begin{array}{ccc}j_{a'}&j_a&L \\ I&I'&J' \end{array}\right\}
\left\{\begin{array}{ccc}j_b&j_{b'}& L \\ I&I'& J \end{array} \right\}.
\label{kkp9.16}
\end{eqnarray}

The final summations that we expect to be able to do in general are over $I$ and $I'$.  Toward this end we need to re express
the $6-j$ symbols that occur in Eq.~(\ref{kkp9.16}) in terms of $6-j$ symbols that depend separately on $I$ and $I'$.  This can be done by the application of Eq.~(6.2.12)
of \cite{Edmonds}.  
We quote the value of $\sum_\chi $ that results from this transformation:
\begin{eqnarray} 
&&\sum_\chi  =\frac{1}{2}\sum_{II'...}F_{aa'b'b}(L)(-1)^{\kappa_a+\kappa_b
+\kappa_{a'}+\kappa_{b'}+j_{a'}+j_b +L+I+I'+I''} \nonumber \\
&&\times(2L+1)(2I''+1)
\left\{\begin{array}{ccc} j_{a'} &j_a & L\\ j_{b'} & j_b & I'' \end{array}\right\}
\left\{\begin{array}{ccc}j_a & I & J'\\J& I'' & j_{b'} \end{array} \right\}
\left\{\begin{array}{ccc}j_{a'}& J' & I'\\J & j_b & I'' \end{array}\right\} \nonumber\\
&&\times (JK-\kappa_{b'}j_{b'}\kappa_{b'}|IK)(J'K-\kappa_a j_a\kappa_a|IK)
\nonumber \\
&&\times(JK-\kappa_b j_b\kappa_b|I'K)(J'K-\kappa_{a'}j_{a'}\kappa_{a'}|I'K)
\nonumber\\
&&\times\chi_{JK-\kappa_{b'}\nu}(j_{b'}\kappa_{b'})
\chi^{\ast}_{JK-\kappa_b\nu}(j_b\kappa_b) \nonumber \\
&&\times\chi_{J'K-\kappa_{a'}\nu'}(j_{a'}\kappa_{a'})
\chi^{\ast}_{J'K-\kappa_a\nu'}(j_a\kappa_a).  
\label{kkp9.17}
\end{eqnarray}

We are finally in a position to do the sums over $I$ and $I'$ by the application of Eq.~(6.2.6) of \cite{Edmonds}.  
This leads to our final {\em exact} result.  It turns out that the 
evaluation of the remaining interaction terms parallels that just described,
differing only in the coupling constants and the single-particle functions that
occur.  For economy of expression, it is this total result that we finally quote:
\begin{eqnarray}
&&\sum = \sum_{I''JJ'\nu\nu'j_a ...\kappa_a ...}
[\frac{1}{2} F_{aa'b'b}(L) \chi_{JK-\kappa_{b'}\nu}(b')
       \chi^{\ast}_{JK-\kappa_b\nu}(b)  \chi_{J'K-\kappa_{a'}\nu'}(a')
       \chi^{\ast}_{J'K-\kappa_a\nu'}(a) \nonumber \\
&&-\frac{1}{2} F_{ba'b'a}(L)
\phi_{JK-\kappa_{b'}\nu}(b')  \phi^{\ast}_{JK-\kappa_b\nu}(b)
 \phi_{J'K-\kappa_{a'}\nu'}(a') \phi^{\ast}_{J'K-\kappa_a\nu'}(a) \nonumber \\
&&+ G_{aa'b'b} \chi_{JK-\kappa_{b'}\nu}(b') \phi^{\ast}_{JK-\kappa_b\nu}(b) 
 \phi_{J'K-\kappa_{a'}\nu'}(a') \chi^{\ast}_{J'K-\kappa_a\nu'}(a)] \nonumber \\
&& \times(-1)^{\kappa_b +\kappa_{b'}-1+j_{a'}+j_b +L+I''} \nonumber \\
&& \times \frac{(2L+1)}{(2J'+1)}\left\{\begin{array}{ccc} j_{a'} & j_a & L\\j_{b'} & j_b & I''\end{array}\right\} \nonumber \\
&&\times(j_a -\kappa_a j_{b'}\kappa_{b'}|I''\kappa_{b'}-\kappa_a)
(j_{a'}-\kappa_{a'}j_b\kappa_b|I''\kappa_b -\kappa_{a'}) \nonumber \\
&&\times (I''\kappa_{b'}-\kappa_a JK-\kappa_{b'}|J'K-\kappa_a)
       (I''\kappa_b -\kappa_{a'} JK-\kappa_b|J'K-\kappa_{a'}) .
\label{kkp9.18}
\end{eqnarray}

We now collect the results of the calculations presented in this section.  The strong coupling limit of the functional ${\cal F}$ for the axial case is given by
the sum of Eqs.~(\ref{kkp9.8a}), (\ref{kkp9.11}), (\ref{kkp9.8b}), 
(\ref{kkp9.11b}), and (\ref{kkp9.18}).  By varying in turn with respect to
$\chi^*_{JK-\kappa_a\nu}(a)$ and $\phi^*_{JK-\kappa_a\nu}$, we obtain the equations of motion
\begin{eqnarray}
\bar{\varepsilon}_{J\nu}\chi_{JK-\kappa_a\nu}(a)&=&  \epsilon_a^{\prime}
\chi_{JK-\kappa_a\nu}(a) -\frac{1}{2}\bar{f}(JK)\sqrt{(J+K-\kappa_a)
(J-K+\kappa_a +1)} \nonumber \\
&&\times\sqrt{(j_a-\kappa_a)(j_a+\kappa_a+1)}\chi_{JK-\kappa_a-1\nu}(a) \nonumber\\
&&+ \sqrt{(J-K+\kappa_a)(J+K-\kappa_a+1)}\sqrt{(j_a+\kappa_a)(j_a-\kappa_a +1)}
\chi_{JK-\kappa_a+1\nu}(a) \nonumber \\
&&+2(K-\kappa_a)\kappa_a\chi_{JK-\kappa_a\nu}(a) \nonumber \\
&&+[F_{aa'b'b}(L)\chi_{J'K-\kappa_{b'}\nu'}(b')\chi^*_{J'K-\kappa_b\nu'}(b)
\chi_{JK-\kappa_{a'}\nu}(a')  \nonumber \\
&&+G_{aa'b'b}(L)\chi_{J'K-\kappa_{b'}\nu'}(b')\phi^*_{J'K-\kappa_b\nu'}(b)
\phi_{JK-\kappa_{a'}\nu}(a')]  \nonumber \\
&&\times\frac{2L+1}{2J+1}\left\{\begin{array}{ccc}j_{a'} &j_a&L \\
j_{b'} &j_b &I \end{array}\right\}  \nonumber \\
&&\times(j_a-\kappa_a j_{b'}\kappa_{b'}|I\kappa_{b'}-\kappa_{a})
(j_{a'}-\kappa_{a'} j_{b}\kappa_{b}|I\kappa_{b}-\kappa_{a'}) \nonumber \\
&&\times(I\kappa_{b'}-\kappa_a J'K-\kappa_{b'}|JK-\kappa_a)
(I\kappa_{b}-\kappa_{a'} J'K-\kappa_{b}|JK-\kappa_{a'}),  \label{PRMa}\\
\underline{\varepsilon}_{J\nu}\phi_{JK-\kappa_a\nu}(a)&=&  -\epsilon_a^{\prime\prime}
\phi_{JK-\kappa_a\nu}(a) -\frac{1}{2}\underline{f}(JK)\sqrt{(J+K-\kappa_a)
(J-K+\kappa_a +1)} \nonumber \\
&&\times\sqrt{(j_a-\kappa_a)(j_a+\kappa_a+1)}\phi_{JK-\kappa_a-1\nu}(a) \nonumber\\
&&+ \sqrt{(J-K+\kappa_a)(J+K-\kappa_a+1)}\sqrt{(j_a+\kappa_a)(j_a-\kappa_a +1)}
\phi_{JK-\kappa_a+1\nu}(a) \nonumber \\
&&+2(K-\kappa_a)\kappa_a\phi_{JK-\kappa_a\nu}(a) \nonumber \\
&&-[F_{ba'b'a}(L)\phi_{J'K-\kappa_{b'}\nu'}(b')\phi^*_{J'K-\kappa_b\nu'}(b)
\phi_{JK-\kappa_{a'}\nu}(a')  \nonumber \\
&&+G_{aa'b'b}(L)\phi_{J'K-\kappa_{b'}\nu'}(b')\chi^*_{J'K-\kappa_b\nu'}(b)
\chi_{JK-\kappa_{a'}\nu}(a')]  \nonumber \\
&&\times\frac{2L+1}{2J+1}\left\{\begin{array}{ccc}j_{a'} &j_a&L \\
j_{b'} &j_b &I \end{array}\right\}  \nonumber \\
&&\times(j_a-\kappa_a j_{b'}\kappa_{b'}|I\kappa_{b'}-\kappa_{a})
(j_{a'}-\kappa_{a'} j_{b}\kappa_{b}|I\kappa_{b}-\kappa_{a'}) \nonumber \\
&&\times(I\kappa_{b'}-\kappa_a J'K-\kappa_{b'}|JK-\kappa_a)
(I\kappa_{b}-\kappa_{a'} J'K-\kappa_{b}|JK-\kappa_{a'}).  \label{PRMb}
\end{eqnarray}
        
We add the normalization conditions for the particle-rotor model 
that follow from Eq.~(\ref{kc.norma}).  We
find
\begin{equation}
\sum_{\kappa_a} [|\chi_{JK-\kappa_a\nu}(a)|^2 
                +|\phi_{JK-\kappa_a\nu}(a)|^2] = 2j_a +1. \label{kc.normpr}
\end{equation}

\section{Alternative derivation and its cranking limit:axial case}

In addition to the PRM, self-consistent or otherwise, we are interested in the cranking theory, valid in the limit in which a single-particle angular momentum
$j_a$ may be neglected compared to the collective angular momentum.  In principle,
we should be able to derive this limit from the form of the theory developed
in Sec.\ II.  However, the interaction terms, as derived, do not provide a
natural pathway to the limit sought.  Therefore we start anew in this section, 
but concentrate on deriving an approximate version of the PRM in which an 
expansion in $(\langle j\rangle/J)$ has been made, the main difference compared to
the previous calculation residing in the treatment of the interaction terms.
We derive an approximate version of the PRM and then introduce the additional
approximation necessary to reach the cranking limit.

For present purposes it is 
convenient to work in coordinate-spin-isospin space, designated by $x$. 
We work with amplitudes that we refer to as coordinate coefficients of 
fractional parentage (CCFP), 
\begin{eqnarray}
V_{JM\nu}(xIM_IK)&=&\langle JM\nu|\hat{\psi}(x)|\overline{IM_IK}\rangle, \label{kkp9.30}\\
U_{JM\nu}(xIM_IK)&=&\langle JM\nu|\hat{\psi}^{\dag}(x)|\underline{IM_IK}\rangle, 
\label{kkp9.30a}
\end{eqnarray}
where $\hat{\psi}(x)$ is the nucleon destruction operator at the 
space-spin-isospin point $x$.  In terms of these amplitudes, we rewrite the 
variational functional ${\cal F}$ of Eq.~(\ref{kc.var1a}) as
\begin{eqnarray}
{\cal F}&=&[\epsilon(xx')-{\cal E}_{J\nu}\delta(x-x')-\bar{E}^*(IK)\delta(x-x')]
V_{JM\nu}(x'IM_I K)V^*_{JM\nu}(xIM_I K) \nonumber \\
&&[-\epsilon(xx')-{\cal E}_{J\nu}\delta(x-x')-\underline{E}^*(IK)\delta(x-x')]
U_{JM\nu}(xIM_I K)U^*_{JM\nu}(x'IM_I K) \nonumber \\
&&+\frac{1}{2}F(xx'x''x''')V_{J'M'\nu'}(x''IM_I K)V^*_{J'M'\nu'}(x'''I'M_{I'} K)
\nonumber \\
&& \times V_{JM\nu}(x'I'M_{I'} K)V^*_{JM\nu}(xIM_I K) \nonumber \\
&&+G(xx'x''x''')V_{J'M'\nu'}(x''IM_I K)U^*_{J'M'\nu'}(x'''I'M_{I'} K)
\nonumber \\
&& \times U_{JM\nu}(x'I'M_{I'} K)V^*_{JM\nu}(xIM_I K) \nonumber \\
&&-\frac{1}{2}F(x'''x'x''x)U_{J'M'\nu'}(x''IM_I K)VU^*_{J'M'\nu'}(x'''I'M_{I'} K)
\nonumber \\
&& \times U_{JM\nu}(x'I'M_{I'} K)U^*_{JM\nu}(xIM_I K). \label{FPRM}
\end{eqnarray}
We have set $\epsilon' =\epsilon'' =\epsilon$ and shall adhere to this simplification for the remainder of our presentation.  To carry out the transformation to Eq.~(\ref{FPRM}), we have made use of a special mode transformation to a basis in which $\epsilon(xx')$ is diagonal,
\begin{eqnarray}
a_\alpha &=& \varphi^*_\alpha(x)\hat{\psi}(x),  \label{mode1} \\
\epsilon_a\varphi_\alpha(x) &=&\epsilon(xx')\varphi_\alpha(x'),  \label{mode2} \\
F(xx'x''x''') &=& F_{\alpha\gamma\delta\beta}\varphi^*_\alpha(x)\varphi_\gamma(x')
\varphi^*_\delta(x'')\varphi_\beta(x''') ,  \label{mode3} \\
G(xx'x''x''') &=& G_{\alpha\gamma\beta\delta}\varphi^*_\delta(x''')\varphi^*_\beta(x')
\varphi_\gamma(x')\varphi_\alpha(x).  \label{mode4} 
\end{eqnarray}

The major device of the present derivation is to transform from angular
momentum eigenfunctions to eigenfunctions localized in angle space,
a technique that has already been exploited in Sec.\ IV.  We base the developments
on expressions for the CCFP that are derived by the same initial transformations that led to Eqs.~(\ref{kkp4.xvee}) and (\ref{kkp4.xyou}), namely
\begin{eqnarray}
\left(\begin{array}{c} V_{JM\nu}(xM_I K) \\U_{JM\nu}(xM_I K) \end{array}\right)
&=&\int dR D^{(J)*}_{MM'}(R)\left(\begin{array}{c}\chi_{JM'\nu}(Rx,K) \\
\phi_{JM'\nu}(Rx,K) \end{array}\right)\sqrt{\frac{2I+1}{8\pi^2}}
D^{(I)}_{M_I K}(R), \label{mode5} \\
\chi_{JM\nu}(Rx,K) &= & \langle JM\nu|\hat{\psi}(Rx)|\hat{0}K\rangle, 
\label{mode6} \\
\phi_{JM\nu}(Rx,K) &= & \langle JM\nu|\hat{\psi}^{\dag}(Rx)|\hat{0}K\rangle, 
\label{mode7} \\
\hat{\psi}(Rx) &=& U^{-1}(R)\hat{\psi}(x)U(R).  \label{mode8}
\end{eqnarray}

When we substitute 
Eq.~(\ref{mode5}) into the first two terms of Eq.~(\ref{FPRM}), we encounter
the restricted completeness relation
\begin{equation}
\sum_{IM} D^{(I)}_{MK}(R)D^{(I)*}_{MK}(R')\frac{2I+1}{8\pi^2} 
= \delta(\hat{n}-\hat{n}')\exp[-iK(\gamma-\gamma')]\frac{1}{2\pi}. \label{kkp9.32b}
\end{equation}
To perform the integral over $\gamma'$, we note the relation (\ref{kc.intr2}), and in the further calculation we utilize basic properties of the $D$ functions,
following from their definition, Eq.~(\ref{kkp4.wigner}), the first one also used extensively for the interaction term to be computed below,
\begin{eqnarray}
\sum_M D^{(J)}_{M'M}(R^ {-1})D^{(J)}_{MM''}(R')&=& D^{(J)}_{M'M''}(R^{-1}R'),
\label{kkp9.37} \\
D^{(J)}_{MM'}(0) &=& \delta_{M,M'}.   \label{kkp9.38}
\end{eqnarray}
Finally, invoking the rotational invariance of the Hamiltonian, which for the term under consideration means that $\epsilon(RxRx')=\epsilon(xx')$, we obtain the result (recall that factors of $8\pi^2$ are suppressed)
\begin{equation}
[\epsilon(xx')-{\cal E}_{J\nu}\delta(x-x')]\chi_{JM\nu}(x')\chi^{\ast}_{JM\nu}(x).  \label{kkp9.39}
\end{equation}
We note the corresponding contributions
\begin{equation}
[-\epsilon(xx')-{\cal E}_{J\nu}\delta(x-x')]
\phi_{JM\nu}(x)\phi^{\ast}_{JM\nu}(x').  \label{kkp9.39a}
\end{equation}

Applying Eqs.~(\ref{kkp9.37}) and (\ref{kkp9.38}) to the interaction term,
we find at the stage that the sums over $IM$ and $I'M'$ have been carried out,
\begin{eqnarray}
&&\frac{1}{2}\int dRdR' F(xx'x''x''')D^{(J)}_{M'M''}(R^{-1}R')
D^{(J')\ast}_{M^{iv}M'''}(R^{-1}R') \nonumber \\
&&\times \chi_{JM'\nu}(Ry')\chi^{\ast}_{JM''\nu}(R'y)
         \chi_{J'M'''\nu'}(R'x')\chi^{\ast}_{J'\nu^{iv}\nu'}(Rx).  \label{kkp9.40}
\end{eqnarray}
Introducing the definition                            
\begin{equation}
R^{-1}R' = {\cal R},   \label{kkp9.41}
\end{equation}
and replacing the integral over $R'$ by an integral ${\cal R}$, we could do
the integrals exactly by decomposing the amplitudes $\chi$ into irreducible tensors.  We have resisted the temptation to do this, since a full calculation
was carried out in Sec.\ III.  It is more illuminating, as well as simpler,
to proceed approximately by expanding ${\cal R}$ about the unit
matrix where ever it appears as the argument of a $\chi$ function.  This
brings in at each order angular momentum operators acting on single-particle
wave-functions and therefore dimensionally is the source of the 
expansion in $(\langle j\rangle/J)$.  For the interaction term the cranking 
limit will arise from the leading term of this expansion.

With the help of the rotational invariance of the interaction, in the present 
instance the relation
for example $F(RxRx'Rx''Rx''')=F(xx'x''x''')$, and the orthonormality relations of the $D$ functions,
we reach the result
\begin{eqnarray}
&& \frac{1}{2}\sum_{JMM'\nu\nu'}\frac{1}{2J+1}F(xx'x''x''')
\chi_{JM\nu'}(x'')\chi^{\ast}_{JM'\nu'}(x''')
                \chi_{JM'\nu}(x')\chi^{\ast}_{JM\nu}(x),
\label{kkp9.43}
\end{eqnarray}
which is almost the cranking limit.  The remaining interaction terms may be
written down by inspection, namely
\begin{eqnarray}
&& \sum_{JMM'\nu\nu'}\frac{1}{2J+1}[G(xx'x''x''')
\chi_{JM\nu'}(x'')\phi^{\ast}_{JM'\nu'}(x''')
                \phi_{JM'\nu}(x')\chi^{\ast}_{JM\nu}(x)  \nonumber \\
&& -\frac{1}{2}F(x'''x'x''x)
\phi_{JM\nu'}(x'')\phi^{\ast}_{JM'\nu'}(x''')
                \phi_{JM'\nu}(x')\phi^{\ast}_{JM\nu}(x)].  \label{kkp9.43a}
\end{eqnarray}

It remains for us to calculate the Coriolis coupling terms.  Remarking that
$\bar{E}^*(IK)$ is an eigenvalue of $D^{(I)}_{MK}$,
\begin{equation}
\bar{E}^*(IK)D^{(I)}_{MK}= \bar{E}^*(\vec{I}^2_{{\rm op}}-K^2)D^{(I)}_{MK},
\label{deig}
\end{equation}
using completeness and integrating by parts, we reach an intermediate stage of the
calculation, in the form,
\begin{eqnarray}
&& -\int dR[\bar{E}^*(\vec{I}^2_{{\rm op}}-K^2)D^{(J)*}_{MM'}(R)\chi_{JM'\nu}
(Rx)] D^{(J)}_{MM'}\chi^*{JM''\nu}(Rx).  \label{kck.24}
\end{eqnarray}
By distribution of the derivatives, $\vec{I}_{{\rm op}}\rightarrow 
\vec{J}_{{\rm op}} +\vec{j}_{{\rm op}}$, by noting the relation
\begin{equation}
\vec{j}_{{\rm op}}(R)\chi(Rx) =-\vec{j}(x)\chi(Rx),   \label{jrev}
\end{equation}
and by expansion in powers of $(j/J)$, Eq.~(\ref{kck.24}) becomes to first
order (in the intrinsic system)
\begin{eqnarray}
&&-\sum_{JM\nu}\bar{E}^*[J(J+1)-K^2]|\chi_{JM\nu}(x)|^2 \nonumber \\
&&+[\frac{\partial \bar{E}^*}{\partial J_i}D^{(J)*}_{MM'}(R)][j_i(x)\chi_{JM'\nu}
(Rx)]D^{(J)}_{MM''}(R)\chi^*_{JM''\nu}(Rx) + ... \label{kck.24a}
\end{eqnarray}
The second term, the Coriolis coupling may be evaluated further by substituting
the general relation (\ref{kkp9.9}) and the matrix elements for intrinsic components,
\begin{eqnarray}
J_-D^{(J)*}_{MM'} &=& \sqrt{(J+M')(J-M'+1)}D^{(J)*}_{MM'-1},  \label{jel1}\\
J_+D^{(J)*}_{MM'} &=& \sqrt{(J-M')(J+M'+1)}D^{(J)*}_{MM'+1},  \label{jel2}
\end{eqnarray}
yielding
\begin{eqnarray}
&&\sum_{JM\nu}\bar{f}(JK)\{\frac{1}{2}\sqrt{(J-M)(J+M+1)}j_+\chi_{JM+1\nu}(x)
\nonumber \\
&&+\frac{1}{2}\sqrt{(J+M)(J-M+1)}j_-\chi_{JM-1\nu}(x)
+Mj_3\chi_{JM\nu}(x)\}\chi^*_{JM\nu}(x)\}.   \label{cor1}
\end{eqnarray}

The corresponding calculation for the other Coriolis term yields the sum
\begin{eqnarray}
&&-\sum_{JM\nu}\underline{E}^*[J(J+1)-K^2]|\phi_{JM\nu}(x)|^2 \nonumber \\
&&+\sum_{JM\nu}\underline{f}(JK)\{\frac{1}{2}\sqrt{(J-M)(J+M+1)}
j_+\phi_{JM+1\nu}(x)
\nonumber \\
&&+\frac{1}{2}\sqrt{(J+M)(J-M+1)}j_-\phi_{JM-1\nu}(x)
+Mj_3\phi_{JM\nu}(x)\}\phi^*_{JM\nu}(x).  \label{cor2}
\end{eqnarray}
The signs in the Coriolis terms (\ref{cor1}) and (\ref{cor2}) appear to be reversed
compared to those encountered in Eqs.~(\ref{kkp9.11}) and (\ref{kkp9.11b}),
but when due account is taken of Eq.~(\ref{jrev}), there is no inconsistency.

We are finally ready to discuss the cranking limit.  The essential observation
is that once the expansion to leading order in $(\langle j\rangle/J)$ has
been made both in the Coriolis coupling and in the interaction terms, the 
resulting approximate functional ${\cal F}$ presents itself as a single sum over $J$.  However, angular momentum is still conserved at this juncture.
We lose angular momentum conservation by assuming that consistent with the
condition $(\langle j\rangle/J)<<1$ we may identify $M$ and $K$, i.\ e., 
we may neglect the angular momentum transferred to or from the particle,
and write, furthermore, ($\omega$ defined below)
\begin{eqnarray}
\chi_{JK\nu}(x)&\rightarrow& \sqrt{2J(\omega)+1}
\chi_{\omega\nu}(x),  \label{kkp9.11d} \\
\chi_{JK\pm 1\nu}(x)&\rightarrow
&\sqrt{2J(\omega)+1}C_{\mp}\chi_{\omega\nu}(x), \label{kkp9.replace}
\end{eqnarray}
i.\ e., the amplitudes differing in $K$ from the ``central value'' by a unit
are assumed proportional to the central amplitude (which is defined as the 
cranking amplitude) up to scale factors $C_{\mp}$ discussed below.  Similar
definitions hold for the $\phi$ amplitudes.  The factor $\sqrt{2J+1}$ is inserted
for convenience, as will be evident from Eq.~(\ref{varcrank}) given below.

These assumptions suggest the following definitions of the components of 
the angular frequency (overline and underline understood)
\begin{eqnarray}
\omega_{\mp}(K)&=&f(JK)C_{\mp}\sqrt{(J\mp K+1)(J\pm K)},  \label{omega1}\\
\omega_3 &=&f(JK)K.   \label{omega2}
\end{eqnarray}
The introduction of the factors $C_{\mp}$ may appear gratuitous at first
sight, but it is needed, as will become especially evident when we treat
the triaxial case, to guarantee that in the cranking limit the theorem that the angular velocity is proportional to the angular momentum is valid \cite{KO}.

Remembering the definition (\ref{defep}) 
and reinstating Cartesian intrinsic coordinates for the Coriolis coupling terms,
we obtain the cranking variational expression
\begin{eqnarray}
[{\cal F}/(2J(\omega) +1)] &=& \epsilon(xx')\chi_{\omega\nu}(x')
\chi^*_{\omega\nu}(x) -\epsilon(xx')\phi_{\omega\nu}(x)\phi^*_{\omega\nu}(x')
\nonumber \\
&&+(\bar{\omega}_i j_i \chi_{\omega\nu}(x))\chi^*_{\omega\nu}(x)
+(\underline{\omega}_i j_i \phi_{\omega\nu}(x))\phi^*_{\omega\nu}(x)
\nonumber \\ 
&& +\frac{1}{2}F(xx'x''x''')\chi_{\omega\nu'}(x'')\chi^*_{\omega\nu'}(x''')
\chi_{\omega\nu}(x')\chi^*_{\omega\nu}(x)  \nonumber \\
&& +G(xx'x''x''')\chi_{\omega\nu'}(x'')\phi^*_{\omega\nu'}(x''')
\chi_{\omega\nu}(x')\phi^*_{\omega\nu}(x)  \nonumber \\
&& -\frac{1}{2}F(x'''x'x''x)\phi_{\omega\nu'}(x'')\phi^*_{\omega\nu'}(x''')
\phi_{\omega\nu}(x')\phi^*_{\omega\nu}(x)  \nonumber \\
&& -\bar{\varepsilon}_{\omega\nu}\chi_{\omega\nu}(x)\chi^*_{\omega\nu}(x)
 -\underline{\varepsilon}_{\omega\nu}\phi_{\omega\nu}(x)\phi^*_{\omega\nu}(x).  \label{varcrank}
\end{eqnarray}
The equations of motion that follow are number-conserving, and according to the 
definitions (\ref{omega1}) and (\ref{omega2}) allow solutions with principal axis cranking

\section{Triaxial rotor: core-particle coupling model and cranking limit}

In this section, we assume that states of interest of neighboring even nuclei can be described phenomenologically by a Hamiltonian
\begin{equation}
{\cal H}_c =\frac{1}{2}a_i I_i^2 +\frac{1}{4}a_{ij}
\{I_i^2,I_j^2\} + ...  \;. \label{kt.1}
\end{equation}
In the calculations to be described below, we shall retain only
the first term of ${\cal H}_c$.  The underlying model arises as
follows: We assume that we can identify states of the appropriate 
even nucleus as $|IM_I n\sigma\rangle$, which we read as the 
$n${\em th} state of angular momentum $I$ belonging to a 
triaxial intrinsic structure $\sigma$.  We also define a rotated
intrinsic state
\begin{equation}
 |R\sigma\rangle = U(R)|\hat{0}\sigma\rangle.  \label{kt.2}
\end{equation}
It is part of the definition of the model that the scalar
product 
\begin{equation}
\langle R\sigma|IM_I n\sigma\rangle \equiv F^{(I)}_{M_I n}(R)
\label{kt.3}
\end{equation}
satisfies the eigenvalue equation
\begin{equation}
{\cal H}_c F^{(I)}_{M_I n} = E^*(In)F^{(I)}_{M_I n}. \label{kt.4}
\end{equation}
Further useful equations satisfied by or defining the model
include
\begin{eqnarray}
|IM_I n\sigma\rangle &=& |IM_I K\sigma\rangle c^{(I\sigma)}_{K n} \label{kt.5} \\
\delta_{nn'} &=& \sum_{K}c^{(I\sigma)*}_{K n}c^{(I\sigma)}_{K n'} \label{kt.6} \\
\delta_{KK'} &=& \sum_n c^{(I\sigma)*}_{Kn}c^{(I\sigma)}_{K'n},
\label{kt.7} \\
\langle R\sigma|IMK\sigma\rangle &=& \sqrt{\frac{2I+1}
{8\pi^2}}D^{(I)}_{MK}(R). \label{kt.8}
\end{eqnarray}

We turn to the evaluation of the terms in the variational
functional ${\cal F}$.  We shall follow the methods of both Sec.\ III and 
Sec.\ IV, depending on the aim of a particular fragment of the calculation.
Starting from the representation
\begin{eqnarray}
V_{JM\nu}(\alpha IM_I n\sigma) &=& \langle JM\nu|a_\alpha|R\sigma\rangle F^{(I)}_{M_I n}(R) \nonumber \\
&=& \langle JM\nu|a_\alpha|R\sigma\rangle D^{(I)}_{M_I K}(R)
c^{(I\sigma)}_{K n},  
\end{eqnarray}
we can derive a formula for the current version of the CFP 
$V$ that is analogous to Eq.~(\ref{kkp4.xvee}), namely
\begin{eqnarray}
V_{JM\nu} (\alpha; IM_I n\sigma) &=& \sum_{K\kappa_a}\sqrt{\frac{8\pi^2}{2j_a +1}}
(-1)^{J-M}   \nonumber \\
&& \times (IM_I J-M|j_a m_a)   
(JK j_a\kappa_a|IK +\kappa_a) \nonumber \\
&& \times(-1)^{j_a +\kappa_a}
\chi_{JK\nu} (j_a\kappa_a\sigma) c^{(I\sigma)}_{K+\kappa_a n}, \label{kc.tcfp1}  \\
(-1)^{j+m}\chi_{JK\nu}(jm\sigma)&=&\langle JK\nu|a_{jm}|\hat{0}
\sigma\rangle.  \label{kc.tdef1}
\end{eqnarray}
The corresponding formula for the CFP $U$ is
\begin{eqnarray}
U_{JM\nu}(\alpha IM_I n\sigma) &=&\sum_{K\kappa_a} \sqrt{\frac{8\pi^2}{2j_a +1}}
\nonumber \\
&& \times (-1)^{J-M +j_a -\kappa_a +j_a +m_a}(IM_I J-M|j_a m_a) \nonumber \\
&&\times(JK j_a \kappa_a|IK+\kappa_a) 
\phi_{JK\nu}(j_a\kappa_a\sigma)c^{(I\sigma)}_{K+\kappa_a n},  
 \label{kc.tcfp2}  \\ 
\phi_{JM\nu} (j_a\kappa_a\sigma) &=& \langle JM\nu|a^{\dag}_{j_a -\kappa_a}|\hat{0}\sigma\rangle.   \label{kc.tdef2} 
\end{eqnarray}

With these formulas, we find the contributions of the simplest
single-particle terms to take the form, in the shell-model or mode representation,
\begin{eqnarray}
\sum_{JK\nu j_a\kappa_a}[(\epsilon_a-\bar{{\cal E}}_{J\nu})|\chi_{JK\nu}(a)|^2
   \nonumber \\
-(\epsilon_a +\underline{{\cal E}}_{J\nu}|\phi_{JK\nu}(a)|^2].  \label{kc.tsp1}
\end{eqnarray}

We study next the term involving the Lagrange multiplier
$\bar{E}^*(In)$.  Here it is convenient to carry out the calculation by a method analogous to that utilized beginning with Eq.~(\ref{deig}). 
With the help of the defining equation
(\ref{kt.4}) and a subsequent integration by parts, we have first of all
\begin{eqnarray}
\bar{E}^*(In)V_{JM\nu}(IM_I n\sigma)&=& 
\int dR [{\cal H}_c(\hat{I}_i)\langle JM\nu|a_\alpha|R\sigma\rangle]
F^{(I\sigma)}_{M_I n}(R).  \label{kt.cor1}
\end{eqnarray}
With the help of the completeness relation
\begin{equation}
\sum_{IM_I n} F^{(I\sigma)}_{M_I n}(R)F^{(I\sigma)*}_{IM_I n}(R')
  =\delta(R-R'),   \label{compl}
\end{equation}
we thus find for the total term 
\begin{equation}
\bar{E}^*(In\sigma)|V_{JM\nu}(\alpha IM_I n\sigma)|^2
= \int dR [{\cal H}_c(\hat{I}_i)\langle JM\nu|a_\alpha|R\sigma\rangle]
JM\nu|a_\alpha|R\sigma\rangle^*. \label{kt.cor2}
\end{equation}
The square bracket may be reexpressed as 
\begin{equation}
{\cal H}_c(\hat{I}_i)\langle JM\nu|a_\alpha|R\sigma\rangle
= [{\cal H}_c(\hat{I}_i)D^{(J)*}_{MK}(R)D^{(j_a)*}_{m_a\kappa_a}(R)]
\langle JK\nu|a_{j_a\kappa_a}|\hat{0}\sigma\rangle. \label{kt.cor3}
\end{equation}
As far as the application of ${\cal H}_c$ in (\ref{kt.cor3})
is concerned, we then write
\begin{eqnarray}
{\cal H}_c(\hat{I}_i) &\rightarrow& {\cal H}_c(\hat{J}_i + \hat{j}_i) \nonumber \\
&=& {\cal H}_c(\hat{J}_i) +\frac{\partial {\cal H}_c}
{\partial \hat{J}_i}\hat{j}_i + ...\;\;, \label{kt.cor4}
\end{eqnarray}
and work only to the order indicated explicitly.

At the same time it is convenient to rewrite
\begin{eqnarray}
{\cal H}_c(\hat{I}_i)&=& \frac{1}{4}b_1(I_+ I_- +I_- I_+)
+\frac{1}{4}b_2(I_+^2 + I_-^2) +\frac{1}{2}b_3 I_3^2
\label{kt.rham} \\
a_1 =b_1 +b_2 ,&& a_2 = b_1 -b_2,\;\; a_3=b_3.
\end{eqnarray}
It is now straightforward  to calculate the contributions arising
from the two terms of Eq.~(\ref{kt.cor4}).  For the first term we find
\begin{eqnarray}
&& - 8\pi^2\{[\frac{1}{2}b_1[J(J+1) -K^2] +\frac{1}{2}b_3K^2]|\chi_{JK\nu}
(a)|^2    \nonumber \\
&& \frac{1}{4}b_2\sqrt{(J-K+2)(J-K+1)(J+K-1)(J+K)}
\chi_{JK-2\nu}(a)\chi^*_{JK\nu}(a)
\nonumber \\
&& +\frac{1}{4}b_2\sqrt{(J+K+2)(J+K+1)(J-K-1)(J-K)} \nonumber \\
&&\times\chi_{JK+2\nu}(a)\chi_{JK\nu}^*(a)\},  \label{kt.cor5}
\end{eqnarray}
and for the second term,
\begin{eqnarray}
&&-8\pi^2 [\frac{1}{2}b_1\sqrt{(J-K+1)(J+K)(j_a +\kappa_a+1)(j_a -\kappa_a)}\chi_{JK-1\nu}(j_a\kappa_a +1\sigma) \nonumber \\
&& +\frac{1}{2}b_1\sqrt{(J+K+1)(J-K)(j_a -\kappa_a+1)(j_a +\kappa_a)}\chi_{JK+1\nu}(j_a\kappa_a-1\sigma) \nonumber \\
&&+\frac{1}{2}b_2\sqrt{(J-K+1)(J+K)(j_a -\kappa_a+1)
(j_a+\kappa_a)}\chi_{JK-1\nu}(j_a\kappa_a-1\sigma) \nonumber \\
&&+\frac{1}{2}b_2\sqrt{(J+K+1)(J-K)(j_a+\kappa_a+1)
(j_a -\kappa_a)}\chi_{JK+1\nu}(j_a\kappa_a+1\sigma) \nonumber \\
&& +b_3 K\kappa_a\chi_{JK\nu}(j_a\kappa_a\sigma)]\chi^*_{JK\nu}
(j_a\kappa_a\sigma).    \label{kt.cor6}
\end{eqnarray}

Both of these terms can be identified as familiar structures.
By means of this identification
we shall have achieved both a simpler form
for the particle-rotor formalism  and for its limiting case, the
cranking formalism.  First consider Eq.~(\ref{kt.cor5}).  Note
that the content of Eqs.~(\ref{kt.4}) and (\ref{kt.5}) can be 
rewritten as
\begin{eqnarray}
{\cal H}_c D^{(I)}_{M_I K} &=& D^{(I)}_{M_I K^{\prime}}
({\cal H}_c)_{K^{\prime}K},   \label{kt.cor7} \\
({\cal H})_{K K^{\prime}}c^{(I\sigma)}_{K^{\prime}n} &=&
E^*(In)c^{(I\sigma)}_{K n}.   \label{kt.cor8}
\end{eqnarray}
This eigenvalue equation was associated with even nuclei and thus with integer values of the angular momentum.  By analytic continuation, we can define a corresponding eigenvalue equation for odd nuclei as follows:  
\begin{equation}
({\cal H}_c(\hat{J}_i))_{KK'}c^{(J)}_{K'\tau} =
E^*(J\tau)c^{(J)}_{K\tau},   \label{kt.cor9}
\end{equation}
where $J,K$ are now half-integral.
We then see that
if we introduce a new set of particle amplitudes $\chi_{J\tau\nu}$
by means of the equation
\begin{equation}
\chi_{JK\nu}(j\kappa) = c^{(J)}_{K\tau}\chi_{J\tau\nu}(j\kappa),
\label{kt.cor10}
\end{equation}
we can transform Eq.~(\ref{kt.cor5}) into the form
\begin{equation}
-E^*(J\tau)|\chi_{J\tau\nu}(a)|^2.   \label{kt.cor11}
\end{equation}
Finally, as we did for the axial case, we can combine energy
terms by means of a definition
\begin{equation}
\varepsilon_{J\nu} ={\cal E}_{J\nu}+ E^*(J\tau). \label{kt.cor12}
\end{equation}

We turn our attention next to Eq.~(\ref{kt.cor6}).  We note first
that this expression is an expanded version of 
\begin{equation}
-[a_i\hat{J}_i\hat{j}_i\chi_{JK\nu}(a)]
\chi^*_{JK\nu}(a),  \label{kt.cor13}
\end{equation}
where $\hat{J}_i$ acts on the value of $K$ and $\hat{j}_i$
acts on the value of $\kappa_a$.  Transforming to the new amplitudes $\chi_{J\tau\nu}$, expression (\ref{kt.cor13}) becomes
\begin{equation}
-[a_i\hat{J}_i\hat{j}_i\chi_{J\tau\nu}(a)]
\chi^*_{J\tau\nu}(a),  \label{kt.cor14}
\end{equation}
where now
\begin{eqnarray}
\hat{J}_i\chi_{J\tau\nu}&=&\chi_{J\tau^{\prime}\nu}(J\tau^{\prime}|J_i|J\tau),        \label{kt.cor15} \\
(J\tau^\prime|\hat{J_i}|J\tau)&=& c^{(J)}_{K'\tau^\prime}
(JK'|J_i|JK)c^{(J)*}_{K\tau}.   \label{kt.cor.16}
\end{eqnarray} 

For the purpose of taking the cranking limit and comparing the forms derived in
Sec.\ IV, we rewrite the results found so far and the corresponding terms
involving $\phi$ amplitudes in coordinate space.  For this we require only
Eqs.~(\ref{kt.cor10}), the corresponding equations
\begin{eqnarray}
\chi_{JK\nu}(x) &=& \langle JK\nu|\hat{\psi}(x)|\hat{0}\sigma\rangle
\label{kt.cor16a}  \\
&=& c^{(J)}_{K\tau}\chi_{J\tau\nu}(x), \label{kt.cor16b}
\end{eqnarray}
and the similar equations for the terms involving $\phi$.
We thus find the contributions
\begin{eqnarray}
&&-\bar{\varepsilon}_{J\nu}|\chi_{J\tau\nu}(x)|^2
+\epsilon(xx')\chi_{J\tau\nu}(x')\chi^*_{J\tau\nu}(x)
+\bar{a}_i [J_i j_i(x)\chi_{J\tau\nu}(x)]\chi^*_{J\tau\nu}(x) \nonumber \\
&&-\underline{\varepsilon}_{J\nu}|\phi_{J\tau\nu}(x)|^2
+\epsilon(xx')\phi_{J\tau\nu}(x')\phi^*_{J\tau\nu}(x)
+\underline{a}_i [J_i j_i(x)\phi_{J\tau\nu}(x)]\phi^*_{J\tau\nu}(x). \label{kt.cor16c} \end{eqnarray}

The cranking limit of these terms may now be taken by means of the replacements
that generalize
Eqs.~(\ref{kkp9.replace}) and (\ref{kkp9.11d}), 
\begin{equation}
\chi_{J\tau'\nu}(x) \rightarrow \sqrt{2J+1}C_\tau(\tau')
\chi_{\omega\nu}(x),    \label{kt.cor17}
\end{equation}
and $\tau'$ refers to $\tau$ or any of the values coupled to
$\tau$ by the matrices of $\hat{J}_i$, with $C_\tau(\tau)=1$.  
This is the essential
blurring of angular momentum conservation that takes us from the 
conserving particle-rotor approximation to the cranking
approximation.  It allows us as well to define the components
of the angular velocity in generalization of Eq.~(\ref{kkp9.11d}).
\begin{eqnarray}
\bar{\omega}_i(\tau) &=& \bar{a}_i\sum_{\tau^{\prime}} C_\tau(\tau')
(J\tau^{\prime}|J_i|J\tau),  \label{kt.cor18} \\
\tau^\prime &=& \tau^\prime(\tau).  \label{kt.cor19}
\end{eqnarray}
As usual, there are corresponding equations for the amplitudes $\phi$.

We may thus replace Eq.~(\ref{kt.cor16c}) by its cranking limit
\begin{eqnarray}
&&(2J+1)[-\bar{\varepsilon}_{\omega\nu}|\chi_{\omega\nu}(x)|^2
+\epsilon(xx')\chi_{\omega\nu}(x')\chi^*_{\omega\nu}(x)
+\bar{\omega}_i  j_i(x)\chi_{\omega\nu}(x)]\chi^*_{\omega\nu}(x) \nonumber \\
&&-\underline{\varepsilon}_{\omega\nu}|\phi_{\omega\nu}(x)|^2
+\epsilon(xx')\phi_{\omega\nu}(x')\phi^*_{\omega\nu}(x)
+\underline{\omega}_i j_i(x)\phi_{\omega\nu}(x)]\phi^*_{\omega\nu}(x), 
\label{kt.cor16d} 
\end{eqnarray}
which is indistinguishable in form from the corresponding terms of 
Eq.~(\ref{varcrank}).

It remains for us to compute the contributions of the interaction terms.  We
consider first an exact calculation analogous to that carried out in 
Sec.\ III, starting from the representations (\ref{kc.tcfp1}) and 
(\ref{kc.tcfp2}) for the CFP in the triaxial case.  It is straightforward to
follow the calculations that begin with Eq.~(\ref{kkp9.13}) and culminate with
the result (\ref{kkp9.18}), as soon as one utilizes the orthonormality relations involving the coefficients 
$c^{(I)}_{Kn}$ at the first step.  The final result is
\begin{eqnarray}
&&\sum = \sum_{I''JJ'KK'\nu\nu'j_a ...\kappa_a ...}
[\frac{1}{2} F_{aa'b'b}(L) \chi_{JK-\kappa_{b'}\nu}(b')
       \chi^{\ast}_{JK'-\kappa_b\nu}(b)  \chi_{J'K'-\kappa_{a'}\nu'}(a')
       \chi^{\ast}_{J'K-\kappa_a\nu'}(a) \nonumber \\
&&-\frac{1}{2} F_{ba'b'a}(L)
\phi_{JK-\kappa_{b'}\nu}(b')  \phi^{\ast}_{JK'-\kappa_b\nu}(b)
 \phi_{J'K'-\kappa_{a'}\nu'}(a') \phi^{\ast}_{J'K-\kappa_a\nu'}(a) \nonumber \\
&&+ G_{aa'b'b} \chi_{JK-\kappa_{b'}\nu}(b') \phi^{\ast}_{JK'-\kappa_b\nu}(b) 
 \phi_{J'K'-\kappa_{a'}\nu'}(a') \chi^{\ast}_{J'K-\kappa_a\nu'}(a)] \nonumber \\
&& \times(-1)^{\kappa_b +\kappa_{b'}-1+j_{a'}+j_b +L+I''} \nonumber \\
&& \times \frac{(2L+1)}{(2J'+1)}\left\{\begin{array}{ccc} j_{a'} & j_a & L\\j_{b'} & j_b & I''\end{array}\right\} \nonumber \\
&&\times(j_a -\kappa_a j_{b'}\kappa_{b'}|I''\kappa_{b'}-\kappa_a)
(j_{a'}-\kappa_{a'}j_b\kappa_b|I''\kappa_b -\kappa_{a'}) \nonumber \\
&&\times (I''\kappa_{b'}-\kappa_a JK-\kappa_{b'}|J'K-\kappa_a)
       (I''\kappa_b -\kappa_{a'} JK'-\kappa_b|J'K'-\kappa_{a'}) .
\label{kkp9.188}
\end{eqnarray}
Superficially, the change compared to Eq.~(\ref{kkp9.18}) is that instead of
a fixed value of $K$, we have a double sum over $K$ and $K'$.  The same expression
holds for a finite number of interacting $K$ bands provided the sums are 
restricted correspondingly.

Finally, we consider the calculation of the interaction term by the method of 
Sec.\ IV, needed to obtain the cranking limit.  Here, in place of Eqs.~(\ref{mode5})-(\ref{mode7}), we utilize
the forms
\begin{eqnarray}
\left(\begin{array}{c} V_{JM\nu}(xM_I n) \\U_{JM\nu}(xM_I n) \end{array}\right)
&=&\int dR D^{(J)*}_{MM'}(R)\left(\begin{array}{c}\chi_{JM'\nu}(Rx,\sigma) \\
\phi_{JM'\nu}(Rx,\sigma) \end{array}\right)\sqrt{\frac{2I+1}{8\pi^2}}
F^{(I)}_{M_I n}(R), \label{mode55} \\
\chi_{JM\nu}(Rx,\sigma) &= & \langle JM\nu|\hat{\psi}(Rx)|\hat{0}\sigma\rangle, 
\label{mode66} \\
\phi_{JM\nu}(Rx,\sigma) &= & \langle JM\nu|\hat{\psi}^{\dag}(Rx)|\hat{0}\sigma\rangle.    
\label{mode77} 
\end{eqnarray}
Once the full completeness relation (\ref{compl}) is utilized instead of the restricted
completeness relation (\ref{kkp9.32b}), the calculation mimics the one carried out
in Sec.\ IV.  In terms of the amplitudes $\chi_{J\tau\nu}$ and $\phi_{J\tau\nu}$,
the result is 
\begin{eqnarray}
&& \sum_{J\tau\tau'\nu\nu'}\frac{1}{2J+1}\{\frac{1}{2}F(xx'x''x''')
\chi_{J\tau\nu'}(x'')\chi^{\ast}_{J\tau'\nu'}(x''')
                \chi_{J\tau'\nu}(x')\chi^{\ast}_{J\tau\nu}(x) \nonumber \\
&& +G(xx'x''x''')\chi_{J\tau\nu'}(x'')\phi^{\ast}_{J\tau'\nu'}(x''')
                \phi_{J\tau'\nu}(x')\chi^{\ast}_{J\tau\nu}(x)  \nonumber \\
&& -\frac{1}{2}F(x'''x'x''x)
\phi_{J\tau\nu'}(x)\phi^{\ast}_{J\tau'\nu'}(x')
                \phi_{J\tau'\nu}(x''')\phi^{\ast}_{J\tau\nu}(x'')\}.  \label{int2}
\end{eqnarray}

The cranking limit of this expression is indistinguishable from the corresponding
terms of Eq.~(\ref{varcrank}) just as was the case for the single-particle
terms (\ref{kt.cor16d}).  Thus the {\em form} of the cranking variational principle for the triaxial is indistinguishable from that for the
axial case and need not be written again.  It is understood, however, that we 
are dealing with full three-dimensional cranking, and that the single-particle
wave functions have suitably modified symmetry.

\section{Summary and discussion}

We have studied the microscopic foundations of the particle-rotor model and of
the cranking model for both axial and triaxial nuclei.  The microscopic model
was chosen in a form in which the interaction is given at the outset as a sum
of multipole and pairing forces.  We carried out the study from the point of view
of the Kerman-Klein method based on the equations of motion for single fermion
operators, and this choice of interaction has the advantage
that the c-number equations of motion for the CFP are formally exact.  These equations of motion and an associated variational principle, worked out
in Sec.\ II, form the basis for the remaining considerations.

[{\footnotesize In the earliest papers on the KK approach \cite{kk:1,CKK}, a
more general shell-model interaction was used in the derivation of c-number equations.  An essential part of the derivation involved the introduction of 
the physical arguments needed to separate this interaction into multipole
and pairing contributions.  As a consequence of the limitations of this 
procedure, the equations of motion found from it are not exact.
Nevertheless, the final equations 
are formally equivalent to those utilized in this paper.  The explanation for
this concordance is that in the approach in this paper, the ``error'' involved in the separation has already been
built into the starting Hamiltonian, as a further compromise, widely accepted,
in the definition of the microscopic theory.}]

As the first application, we derived in Sec.\ III a self-consistent 
particle-rotor model for axially symmetric nuclei.  The derivation was carried
out using basic ideas developed in \cite{pav4}, where, starting from a semi-microscopic version of the theory, we derived the standard non-self-consistent
version of the particle-rotor model. The present discussion complements the previous one in the sense that it starts from the beginning and carries the
reasoning up to the edge of the semi-microscopic form.

We began this work with the prejudice that, as apposed to previous treatments, a natural path to the cranking model involved passing through the particle-rotor model.  Though we were ultimately able to confirm this prejudice,  the version 
of the particle-rotor model derived in Sec.\ III, though a useful one for 
applications \cite{pav4}, does not appear to be useful in the further transition to cranking.  For this purpose we must be able to expand all contributions in
powers of $(\langle j\rangle/J)$, the ratio of a characteristic single-particle
angular momentum to the collective angular momentum.  We have not discovered such
an expansion for the interaction forms derived in this first treatment.
Therefore, in Sec.\ IV we start anew, utilizing an approach already described
briefly for two-dimensional rotations in an early publication \cite{KD}.  Rather
than pushing through to a formally exact result, we stop the calculation
at the leading order of the small parameter, and thus obtain an approximate version of the particle-rotor model that still conserves angular momentum,
but is only a step away from the cranking limit.  This further step
violates angular momentum conservation by the way in which an angular velocity is
introduced to replace the collective angular momentum.  In Sec.\ V the considerations of both previous sections are generalized to the triaxial case.

Several special features of our treatment should be highlighted.  For the axial
case, as soon as the neighboring even nuclei are represented by bands with
non-vanishing $K$ values, we have tilted cranking in its simplest form.  
{\em A fortiori}, in the triaxial case we derive the possibility of full
three-dimensional cranking.  Within our mode of analysis, these statements
may be taken to have the status of theorems.  Another feature of our derivations
of cranking models is that number conservation is maintained.

Nevertheless, in the light of recent developments associated with tilted
cranking \cite{F1,F2,F3,F4,F5}, possible limitations on our work have to be
addressed.  Superficially, our results apply to one quasiparticle spectra
of odd nuclei, whereas the current focus of interest is on at least two
quasiparticle spectra of even nuclei, and even more on multi-quasiparticle
states.  In principle, however, these examples are covered by our considerations.
Thus the two quasiparticle case is readily derived from the formalism developed in
Appendix A.  The multiple quasiparticle case is covered if one replaces the
reference ground states of the even nuclei by suitably chosen band heads of
two quasiparticle bands.  Details of such calculations are best addressed
within the framework of specific applications.

\appendix

\section{Further development of the theory: Iterative solution schemes }

The theory developed in Sec.\ II was sufficient for the purposes of the remainder
of the body of this work, a study of the strong-coupling limit.
However, we cannot resist the temptation to show how this theory
 can be developed further, and the full architecture
used to suggest algorithms for the solution of the non-linear problem thus
defined.  We point out before proceeding, moreover,
that applications of algorithms similar to some of those to be described were carried out in our early work \cite{DP,Dreiss}.

To begin the extended development, it is helpful to introduce a more concise
representation of the equations of motion (\ref{kc.eom3}) and (\ref{kc.eom4}) for the
CFP by defining the vector
\begin{equation}
\Psi_i(\alpha n) = \left(\begin{array}{c} V_i(\alpha n) \\ U_i(\alpha n)
\end{array}\right).    \label{kc.psivec}
\end{equation}
The equations of motion can then be written
\begin{equation}
{\cal E}_i \Psi_i(\alpha n) = {\cal H}(\alpha n,\beta n')\Psi_i(\beta n'),
\label{kc.psieq}
\end{equation}
where ${\cal H}(\alpha n,\beta n')$ is the Hermitian matrix
\begin{eqnarray}
&&{\cal H}(\alpha n,\beta n')= \nonumber \\
&&\left(\begin{array}{cc} (\epsilon_a^{\prime}-
E_{\bar{n}}^*)\delta_{nn'}\delta_{\alpha\beta} +\bar{\Gamma}(\alpha n,\beta n')&
\Delta(\alpha n,\beta n') \\  \Delta^{\ast}(\beta n',\alpha n) & (-\epsilon_a^{\prime\prime}-E_{\underline{n}}^*)\delta_{nn'}\delta_{\alpha\beta}+\underline{\Gamma}(\alpha n,\beta n')\end{array}\right),   \label{kc.hamarray}
\end{eqnarray}
and the potentials are defined as
\begin{eqnarray}
\bar{\Gamma}(\alpha n,\gamma n') &=& F_{\alpha\gamma\delta\beta}[V_i^*(\beta n')
V_i(\delta n)] \nonumber \\
&=& F_{\alpha\gamma\delta\beta}{\cal R}_{11}(\delta n,\beta n'), \label{kc.defg1}\\
\underline{\Gamma}(\alpha n,\gamma n')&=&F_{\bar{\delta}\bar{\beta}\bar{\alpha}
\bar{\gamma}}[U_i^*(\beta n')U_i(\delta n)] \nonumber \\
&=& F_{\bar{\delta}\bar{\beta}\bar{\alpha}\bar{\gamma}}{\rm tr}\frac{1}{2}
(1-\tau_3){\cal R}_{22}(\delta n,\beta n'), \label{kc.defg2}\\
\Delta(\alpha n,\gamma n') &=&G_{\alpha\bar{\gamma}\beta\bar{\delta}}
[U_i^*(\delta n')V_i(\beta n)]\nonumber \\
&=&G_{\alpha\bar{\gamma}\beta\bar{\delta}}
{\cal R}_{12}(\beta n,\delta n').
  \label{kc.defg} 
\end{eqnarray}
Here we have utilized a generalized density matrix, 
${\cal R}(\alpha n,\beta n')$,
defined as
\begin{equation}
{\cal R}(\alpha n, \beta n')=\Psi_i(\alpha n)\Psi_i^*(\beta n')=\left(\begin{array}{cc}{\cal R}_{11} &{\cal R}_{12} \\{\cal R}_{21} &
{\cal R}_{22}\end{array}\right),  \label{kc.defR}
\end{equation}
satisfying the idempotent condition
\begin{equation}
{\cal R}^2 = \Omega{\cal R}. \label{kc.idem}
\end{equation}

The structure displayed in the preceeding paragraph suggests that if we knew the potentials $\bar{\Gamma},\underline{\Gamma},\Delta$, the chemical potential
$\lambda$, and the excitation energies $E_{\bar{n}}^*,E_{\underline{n}}^*$,
we could view Eq.~(\ref{kc.psieq}) as a linear eigenvalue problem, with given
Hamiltonian ${\cal H}$, with eigenvalues ${\cal E}_i$ and with solutions 
normalized according to Eq.~(\ref{kc.norm}).  In turn,
this suggests at least the first elements of an iterative scheme for the solution
of Eq.~(\ref{kc.psieq}), where given the $\nu th$ approximation, ${\cal H}^{(\nu)}$
to ${\cal H}$, we define the $(\nu +1)st$ approximation as the solution of the linear eigenvalue equation
\begin{equation}
{\cal E}_i^{(\nu +1)}\Psi_i^{(\nu +1)}= {\cal H}^{(\nu)}\Psi_i^{(\nu +1)}.
\label{kc.it}
\end{equation}
Thus, with the help of Eq.~(\ref{kc.defR}), we can construct a (tentative, see below) $(\nu +1)st$ approximation,
${\cal R}^{(\nu +1)}$, to the density matrix.  Consequently, we can calculate a next approximation to the ``potentials''.

But how do we generate higher approximations to the energies, including
$\lambda$, of the neighboring even nuclei
also needed for the next approximation to ${\cal H}$?  Actually the 
formalism provides several alternatives on how to proceed at this point.
We describe first a method
which adds the minimum number of conditions sufficient to determine the energies.
This is to calculate, e.\ g.,
\begin{eqnarray}
E_{\bar{n}} &=& \langle\bar{n}|H|\bar{n}\rangle \approx E_{\bar{n}}({\cal R}^{(\nu +1)})= E_{\bar{n}}^{(\nu +1)} ,    \label{kc.encal1} \\
E_{\underline{n}} &=& \langle\underline{n}|H|\underline{n}\rangle \approx E_{\underline{n}}({\cal R}^{(\nu +1)})= E_{\underline{n}}^{(\nu +1)} ,    \label{kc.encal2} 
\end{eqnarray}
from which we can form the averages and differences needed for the calculation
of ${\cal H}^{(\nu +1)}$.

Before we can actually proceed to the next iteration, we must take another step
first recognized in our early work on the pairing problem \cite{DP}.  The general theory requires that if we are to interpret the diagonal elements of $H$ as
eigenvalues, then the off-diagonal elements of $H$ must vanish.  The latter may be calculated at any stage of approximations by the same techniques used for the diagonal elements.  In general,
we shall find at any intermediate stage of the calculation that these conditions 
are not fully satisfied.  To rectify this deficiency, we must therefore carry out orthogonal transformations in the spaces $|\bar{n}\rangle$ and $|\underline{n}\rangle$ in order to eliminate the off-diagonal elements.  These
are equivalent to (different) linear transformations of the CFP $V$ and $U$.
Such transformations are norm preserving.

We have now (almost) defined a cycle of the present algorithm.  To achieve a full
measure of self-consistency requires yet an additional link in 
the chain of reasoning.  This is because to this point, the vectors
$\Psi_i$ cannot be guaranteed to satisfy the normalization conditions (\ref{kc.norma}).  In general we shall find
\begin{equation}
\sum_{m_a n} \Psi_i^{\dag}(\alpha n)\Psi_i(\alpha n) =
\Lambda_a \Omega_a,   \label{kc.norml}
\end{equation}
whereas full self-consistency requires $\Lambda_a =1$.  We can rectify the deficiency by rescaling the solution
\begin{equation}
\Psi_i^{(\nu)}(\alpha n) \rightarrow \sqrt{\Lambda_a}\Psi_i^{(\nu)}(\alpha n),  \label{kc.scale}
\end{equation}
redefining the potentials, e.\ g.,
\begin{equation}
\bar{\Gamma}(\alpha n,\gamma n')=F_{\alpha\gamma\delta\beta}
\sqrt{\Lambda_b\Lambda_d}[V_i^{*(\nu)}(\beta n')V_i^{(\nu)}
(\delta n)], \label{kc.spot}
\end{equation}
and similarly for other potentials.  This requires us to fit in an additional 
set of iterations until the values $\Lambda_a =1$ are achieved.

We next describe an alternative algorithm in which the equations of motion
and normalization conditions are used in the same way as in the algorithm just described.  However the transition from ${\cal H}^{(\nu)}$ to ${\cal H}^{(\nu +1)}$ is done differently and requires further development of the formalism.
The first step is to combine the equations of motion (\ref{kc.psieq}) with their
complex conjugate equations so as to eliminate the eigenvalues ${\cal E}_i$.
We thereby obtain the following equations for the generalized density matrix
${\cal R}$,
\begin{equation}
0={\cal R}(\alpha n,\gamma n''){\cal H}(\gamma n'',\beta n')-
{\cal H}(\alpha n,\gamma n''){\cal R}(\gamma n'',\beta n'),  \label{kc.dmeq}
\end{equation}
i.\ e., we find the vanishing of the commutator,
$[{\cal R},{\cal H}]=0$.

Before undertaking the exposition of the algorithm, we exhibit an alternative
derivation of Eq.~(\ref{kc.dmeq}) utilizing a variant of the variational principle
(\ref{kc.var1}), (\ref{kc.var1a}).  Consider the functional
\begin{equation}
{\cal D} ={\cal G}-\Theta(\alpha n,\beta n')[{\cal R}^2(\beta n',\alpha n)
-\Omega{\cal R}(\beta n',\alpha n)],   \label{kc.vardm}
\end{equation}
where the ``new'' constraint with Lagrange multiplier matrix $\Theta$ is for
normalization in the density matrix form.  Since
\begin{equation}
\frac{\delta{\cal G}}{\delta{\cal R}(\beta n',\alpha n)} =
{\cal H}(\alpha n,\beta n'),  \label{kc.varcond}
\end{equation}
it follows that the variational condition applied to Eq.~(\ref{kc.vardm})
yields the equation
\begin{equation}
{\cal H}-\Theta{\cal R}-{\cal R}\Theta +\Theta\Omega =0.  \label{kc.varres}
\end{equation}
From this condition, Eq.~(\ref{kc.dmeq}) is readily derived by forming the appropriate commutator.   

However, Eq.~(\ref{kc.varres}) contains additional
information that we shall exploit below.
Indeed, we shall consider the possibility of constructing an algorithm on the 
basis of Eq.~(\ref{kc.varres}), but first we describe one that utilizes
Eq.~(\ref{kc.dmeq}) for the density matrix.  Recalling the first algorithm
described, let us imagine ourselves at the point where we have an approximation to the density matrix ${\cal R}$ that has been determined from the iterative procedure associated with the equations of motion (\ref{kc.psieq}), and that there
remains the problem of computing the next approximation to the energies of the even neighbors.  An alternative to the procedure that starts with Eq.~(\ref{kc.encal1}) is to note that the equation of motion in the density  
matrix form (\ref{kc.dmeq}) provides a sufficient set of equations to determine
the excitation energies when supplemented by the number conservation conditions
\begin{eqnarray}
0&=& \sum_{\alpha i} |V_i(\alpha n)|^2 - \bar{N}  \nonumber \\
 &=& \sum_\alpha {\cal R}_{11}(\alpha n,\alpha n) -\bar{N} \label{kc.numov} \\
0&=& \sum_{\alpha i} |U_i(\alpha n)|^2 - \Omega +\underline{N}  \nonumber \\
 &=& \sum_\alpha {\cal R}_{22}(\alpha n,\alpha n) -\Omega +\underline{N} \label{kc.numov1} 
\end{eqnarray}
Just as we did previously for the Hamiltonian, we should check that the number
operator matrices $\sum_\alpha{\cal R}_{11}(\alpha n,\alpha n')$ and
$\sum_\alpha{\cal R}_{22}(\alpha n,\alpha n')$ are diagonal and, if necessary,
diagonalize them so as to improve the convergence of the procedure.

So far we have suggested two possible algorithms which used the equations of motion (\ref{kc.psieq}), attached in each case to a different method for 
calculating the energies of the even nuclei.  We next propose two additional
algorithms, also distinguished by one of the two methods of computing the energy
in which, however, we replace the equations of motion for the $\Psi$ vectors
by a condition related to the density-matrix formulation.  Starting from 
Eq.~(\ref{kc.varres}) we derive the pair
\begin{eqnarray}
{\cal R}{\cal H} &=& {\cal R}\Theta{\cal R} ={\cal H}{\cal R}, 
\end{eqnarray}
which imply that 
\begin{equation}
{\cal H} =\Theta\Omega.
\end{equation}
Substituting this result into Eq.~(\ref{kc.varres}),  we obtain
\begin{equation}
{\cal H} - \frac{1}{\Omega}\{{\cal R},{\cal H}\} =0.  \label{kc.rvec}
\end{equation}

We describe an algorithm based on Eq.~(\ref{kc.rvec}) and the normalization condition
${\cal R}^2=\Omega{\cal R}$.  Suppose that in a $\nu th$ approximation, we have
\begin{eqnarray}
({\cal R}^{(\nu))})^2 -\Omega{\cal R}^{(\nu)} &=& 0,  \label{kc.nunorm} \\
{\cal H}^{(\nu)}-\frac{1}{\Omega}\{{\cal R}^{(\nu)},{\cal H}^{(\nu)}\}&=& 
{\cal Y}^{(\nu)}.  \label{kc.nudm}
\end{eqnarray}
As an illustration of the method of steepest descents, we now choose, in order to improve the value of the density matrix
\begin{equation}
\delta{\cal R}^{(\nu)} =-\eta{\cal Y}^{(\nu)},   \label{kc.deltar}
\end{equation}
where $\eta$ is an arbitrary small parameter that may be set to unity for
${\cal Y}^{(\nu)}$ sufficiently small.  We can check that this choice preserves
the norm to first order, since it satisfies the required condition
\begin{equation}
\{{\cal R}^{(\nu)},\delta{\cal R}^{(\nu)}\} -\Omega\delta{\cal R}^{(\nu)}=0,
\label{kc.varnorm}
\end{equation}
that follows from Eq.~(\ref{kc.nunorm}).
To see this, note that Eq.~(\ref{kc.varnorm}) implies that 
\begin{equation}
{\cal R}^{(\nu)}\delta{\cal R}^{(\nu)}{\cal R}^{(\nu)} =0,
\label{kc.projnorm}
\end{equation}
which besides the trivial solution, is satisfied by Eq.~(\ref{kc.deltar}),
as follows from Eq.~(\ref{kc.nudm}).

This as far as we can go without attacking specific models.


\end{document}